\documentclass[prd,aps,preprint,showpacs,nofootinbib,eqsecnum,superscriptaddress]{revtex4-1}

\usepackage{hyperref}
\usepackage{amsfonts,amssymb,amsmath,amsthm,amsxtra,mathtools,euscript,mathrsfs}

\usepackage{bm}
\usepackage{graphicx,color,xcolor,subfigure}
\usepackage{dcolumn}
\usepackage{multirow}

\def\a{\alpha}

\def\s{\sigma}
\def\k{\kappa}

\def\vt{\vartheta}
\def\bvt{\bar{\vartheta}}

\def\vphi{\varphi}
\def\k{\kappa}
\def\k5{\kappa_5}

\newcommand{\tabincell}[2]{\begin{tabular}{@{}#1@{}}#2\end{tabular}}
\newcommand{\ie}{\textit{i}.\textit{e}.}

\newcommand{\etal}{\textit{et al}.}

\newcommand{\lagrangian}{\mathcal{L}}

\def\hzero{\hat{0}}
\def\hone{\hat{1}}
\def\htwo{\hat{2}}
\def\hthree{\hat{3}}

\def\hfive{\hat{5}}

\definecolor{hellmagenta}{rgb}{1,0.75,0.9} 
\definecolor{hellcyan}{rgb}{0.75,1,0.9} 
\definecolor{hellgelb}{rgb}{1,1,0.8} 
\definecolor{colKeys}{rgb}{0,0,1} 
\definecolor{colIdentifier}{rgb}{0,0,0} 
\definecolor{colComments}{rgb}{1,0,0} 
\definecolor{colString}{rgb}{0,0.5,0} 
\definecolor{darkyellow}{rgb}{1,0.9,0}

\begin{document}
\title{Teleparallel Conformal Invariant Models Induced by Kaluza-Klein Reduction}

\author{Chao-Qiang Geng}
\email[Electronic address: ]{geng@phys.nthu.edu.tw}
\affiliation{Chongqing University
of Posts \& Telecommunications, Chongqing, 400065, China}
\affiliation{Department of Physics,
National Tsing Hua University, Hsinchu 300, Taiwan}
\affiliation{Physics Division,
National Center for Theoretical Sciences, Hsinchu 300, Taiwan}
\affiliation{Synergetic Innovation Center for Quantum Effects and Applications (SICQEA),\\
 Hunan Normal University, Changsha 410081, China}
\author{Ling-Wei Luo}
\email[Electronic address: ]{lwluo@mx.nthu.edu.tw}
\affiliation{Department of Physics,
National Tsing Hua University, Hsinchu 300, Taiwan}

\begin{abstract}
We study the extensions of teleparallism in the Kaluza-Klein (KK) 
scenario by writing the analogous form to the torsion scalar $T_{\text{NGR}}$ in terms of the corresponding
antisymmetric tensors, given by
$T_{\text{NGR}} = a\,T_{ijk} \, T^{ijk} 
    + b\,T_{ijk} \,T^{kji} 
    + c\,T^{j}{}_{ji} \, T^{k}{}_{k}{}^{i}$,
in the four-dimensional New General Relativity (NGR) with  arbitrary coefficients 
$a$, $b$ and $c$. After the KK dimensional reduction, 
the Lagrangian in the Einstein-frame
can be realized by taking $2a+b+c=0$ 
with the ghost-free condition $c\leq0$ 
for the one-parameter family of teleparallelism. 
We demonstrate that the pure conformal invariant gravity models can be constructed
by the requirements of $2a+b=0$ and $c=0$. 
In particular, the torsion vector can be identified as the conformal gauge field, 
while the conformal gauge theory can be obtained by $2a+b+4c=0$ or $2a+b=0$,
which is described on the Weyl-Cartan geometry $Y_4$ 
with the ghost-free conditions $2a+b+c>0$ and $c\neq0$.
We also consider the weak field approximation and 
discuss the non-minimal coupled term of the scalar current and torsion vector.
For the  conformal invariant models with $2a+b=0$, 
we find that only the anti-symmetric tensor field is allowed 
rather than the symmetric  one.
\end{abstract}


\maketitle

\section{Introduction}

The well-known five-dimension gravity theory is called 
the Kaluza-Klein (KK) theory~\cite{Kaluza,Klein} with
the concept of unifying electromagnetic
and gravitational forces in the high dimensional spacetime.
The fifth dimension is regarded as a circle $S^1$ under the spontaneous compactification, 
where the full spacetime manifold is a product manifold $M^4 \times S^1$.
The scalar-tensor theories can be induced by 
the KK reduction of the effective theories~\cite{Bailin:1987jd,Overduin:1998pn}.

The alternative gravitational theory under the 
absolutely parallelism was originally proposed by Einstein~\cite{Einstein:ap},
in which the connection is constructed by the commutation coefficients and
the Lagrangian would lead to the symmetric equations of motion~\cite{Einstein:teleaction}.
Nowadays, we recognize that the geometry is the Weitzenb\"{o}ck geometry $T_4$ 
and the corresponding gravitational theory is called 
teleparallel equivalent to general relativity (TEGR).
New General Relativity (NGR) was presented by Hayashi and Shirafuji~\cite{Hayashi:1979qx},
which is one of the generalizations in teleparallelism
with the equivalence outcome as general relativity (GR).
In  NGR, 
the \emph{torsion scalar} is given by
\begin{equation}\label{E:torsion scalar in coordinate frame}
T_{\text{NGR}} = a\,T_{\rho\mu\nu} \, T^{\rho\mu\nu} 
    + b\,T_{\rho\mu\nu} \,T^{\nu\mu\rho} 
    + c\,T^{\nu}{}_{\nu\mu} \, T^{\s}{}_{\s}{}^{\mu},
\end{equation}
where the torsion tensor is defined by
the Weitzenb\"{o}ck connection  and 
the coefficients $a$, $b$ and $c$ are arbitrary.
In particular, the torsion scalar denoted by $T$ is given by 
$(a,b,c)=(1/4,1/2,-1)$ for TEGR.
The choices of the coefficients have been explored 
in the literature~\cite{Hayashi:1979qx,Scherrer:1955,Cho:1975dh,
Hehl:1978yt,Nitsch:1979qn,Maluf:2011kf}.
Recently, several specific extensions in teleparallelism
have been investigated, such as models with 
the teleparallel dark energy~\cite{Geng11} and
$f(T)$~\cite{Ferraro:2006jd, Linder:2010py}.

Teleparallelism in the KK scenario has been
discussed in \cite{deAndrade:1999vq, Barbosa:2002mg,
Fiorini:2013hva, Bamba:2013fta, Geng:2014yya, Geng:2014nfa}. 
In this study, we consider  the high dimensional gravity theory based on the
analogous form to  (\ref{E:torsion scalar in coordinate frame}) in the four dimensional
NGR theory.
It is known that (\ref{E:torsion scalar in coordinate frame}) in
NGR provides  a much richer structure than those in  TEGR and GR
due to the arbitrariness of the coefficients.
Under the KK reduction, the effective Lagrangian will contain
new couplings associated with some combinations of the coefficients.
We will build the conformal invariant models in teleparallelism and
examine the constraints on these new couplings

This paper is organized as follows. 
In Sec.~II, we first introduce  the NGR gravity theory on the Weitzenb\"{o}ck geometry in four dimensions
and then extend  it to a higher dimension.
By following the KK reduction procedure,
we obtain the effective Lagrangian in the four-dimensional spacetime.
In Sec.~III, we present some specific models
with the conformal transformations for the zero modes in the effective Lagrangians.
We study the conditions for having the theory in the Einstein-frame
and with the conformal invariant, respectively.
In Sec.~IV, we analyze the weak field approximation of 
the four-dimensional effective theory and discuss 
the cases in the conformal invariant models.
Finally, we give the conclusions in Sec.~V.

\section{Kaluza-Klein Reduction in Teleparallelism }

\subsection{The Weitzenb\"{o}ck Background}
In the four-dimensional Weitzenb\"{o}ck geometry $T_4$, 
the orthonormal coframe $\vartheta^i = e^i_{\mu} \,dx^{\mu}$ is associated with 
the Minkowski metric $\eta_{ij}=\text{diag}(+1,-1,-1,-1)$, 
which gives the relation of the four-dimensional metric 
$g_{\mu\nu}=\eta_{ij}\, e^i_{\mu}\, e^j_{\nu}$,
where Greek indices $\mu,\nu,\rho,\ldots=0,1,2,3$ 
and Latin indices $i,j,k,\ldots = \hzero,\hone,\htwo,\hthree$ 
denote as the coordinate and orthonormal frame indices, 
respectively.\footnote{We use the hatted number to represent 
the components in the orthonormal frame.}
The frame fields (vielbein) $e^i_\mu$ and their inverse $e_i^\mu$ 
connect the coordinate and orthonormal frames
via $e_i=e^{\mu}_i\,\partial_{\mu}$ and $\vt^i = e^{i}_{\mu}\,dx^{\mu}$, 
which satisfy the relations
$e^i_{\mu}\,e^{\mu}_j=\delta^i_j$ 
and 
$e^{\mu}_i\,e^i_{\nu}=\delta^{\mu}_{\nu}$.
In the Weitzenb\"{o}ck geometry, 
the absolute parallelism of the frame fields
define the Weitzenb\"{o}ck connection from 
the zero covariant derivative of the frame fields
$\nabla_{\nu}e^i_\mu=\partial_{\nu}e^i_\mu-e^i_{\rho}\, \Gamma^{\rho}_{\mu\nu}=0$,
resulting in
$\Gamma^{\rho}_{\mu\nu}= e^{\rho}_i\partial_{\nu}e^i_{\mu}$.
The torsion tensor is given by the anti-symmetric part 
of the Weitzenb\"{o}ck connection
$T^{\rho}{}_{\mu\nu}=e^{\rho}_i(\partial_{\mu}e^i_{\nu}-\partial_{\nu}e^i_{\mu})$.
Consequently, the \emph{torsion scalar}  for TEGR can be 
written as the form in (\ref{E:torsion scalar in coordinate frame})
in the coordinate frame, or
\begin{equation}\label{E:torsion scalar in orthonormal frame}
T = a\,T_{ijk} \, T^{ijk} 
    + b\,T_{ijk} \,T^{kji} 
    + c\,T^{j}{}_{ji} \, T^{k}{}_{k}{}^{i}
\end{equation}
in the orthonormal frame with $(a,b,c)=(1/4,1/2,-1)$.
Subsequently, the torsion scalar can be rewritten as
\begin{equation}
T = \frac{1}{2}\,T^{\rho}{}_{\mu\nu}\,S_{\rho}{}^{\mu\nu}\,,
\end{equation}
where $S_{\rho}{}^{\mu\nu} = K^{\mu\nu}{}_{\rho} +
\delta_{\rho}^{\mu}\,T^{\s\nu}{}_{\s} -
\delta_{\rho}^{\nu}\,T^{\s\mu}{}_{\s}$
with the contorsion
$K^{\rho}{}_{\mu\nu} = (-1/2)(T^{\rho}{}_{\mu\nu} -
T_{\mu}{}^{\rho}{}_{\nu} - T_{\nu}{}^{\rho}{}_{\mu})$.
The integration of the torsion scalar
\begin{equation}\label{E:telaparallel Lagrangian}
S_{\text{TEGR}}= \frac{1}{2\kappa} \int d^4x\, e\, T
\end{equation}
yields the gravitational action equivalent to GR up to the divergence,
so-called TEGR, where $\kappa= 8\pi\,G$ is the gravitational coupling and
$e\equiv \det(e^i_{\mu})$.

The generalization of TEGR can be realized 
by using the arbitrary coefficients $(a,b,c)$
instead of $(1/4,1/2,-1)$ in (\ref{E:torsion scalar in coordinate frame}) and
(\ref{E:torsion scalar in orthonormal frame}).
In NGR, the Lagrangian density in the orthonormal frame is given by
(\ref{E:torsion scalar in orthonormal frame})
without specific values for $(a,b,c)$,
which can be reformulated to be
\begin{equation}\label{E:NGR Lagrangian}
T_{\text{NGR}}  = \frac{1}{2}\,T^{i}{}_{jk}\,\Sigma_{i}{}^{jk}\,,
\end{equation}
where
\begin{equation}\label{E:Sigma tensor}
\Sigma_{i}{}^{jk} 
= 2a\,T_{i}{}^{jk} 
  + b\,(T^{kj}{}_{i} - T^{jk}{}_{i})
  + c\,(\delta^{k}_{i}\,T^{lj}{}_{l} 
        - \delta^{j}_{i}\,T^{lk}{}_{l})\,.
\end{equation}

\subsection{The Kaluza-Klein Reduction}

\textbf{(i) Geometric part}

In the five-dimensional Weitzenb\"{o}ck geometry $T_5$, the metric is given by
$\bar{g}_{MN}=\bar{\eta}_{IJ}\, e^I_{M}\, e^J_{N}$,
where $M,N=0,1,2,3,5$, $I,J=\hzero,\hone,\htwo,\hthree,\hfive$,
and $\bar{\eta}_{IJ}=\text{diag}(+1,-1,-1,-1,-1)$
is the metric in the orthonormal frame.
In the orthonormal frame, it is convenient to calculate torsion components
by the first Cartan structure equation $\bar{T}^{I}=d\bvt^{I}$.
The decompositions of the torsion 2-forms are
$\bar{T}^{i}=(1/2)\,\bar{T}^{i}{}_{jk}\,\bvt^{j}\wedge\bvt^{k}
+\bar{T}^{i}{}_{\hfive j}\,\bvt^{\hfive}\wedge\bvt^{j}$ and 
$\bar{T}^{\hfive}=(1/2)\,\bar{T}^{\hfive}{}_{ij}\,\bvt^{i}\wedge\bvt^{j}
+\bar{T}^{\hfive}{}_{\hfive i}\,\bvt^{\hfive}\wedge\bvt^{i}$.

In the Kaluza-Klein theory, we take the five-dimensional spacetime
to be a product manifold $T_5=T_4\times S^1$ 
with the fifth dimension compactifying to $S^1$
and the five-dimensional metric $\bar{g}$ in the coordinate frame as
\begin{equation}\label{E:KK metric}
\bar{g}_{MN} =
\begin{pmatrix}
    g_{\mu\nu}-\kappa^2\phi^2A_{\mu}A_{\nu}  &  \kappa\,\phi^2A_{\mu}  \\
    \kappa\,\phi^2A_{\nu}                             &  -\phi^2
\end{pmatrix}\,,
\end{equation}
where $\bar{g}_{\mu\nu}$, $A_{\mu}$ and $\phi$ depend on 
$x^{\mu}$ and $y$ coordinates, 
and the constant $\kappa$ has a mass dimension of $[\kappa]=-1$.
The corresponding coframes in $T_5$ are
\begin{subequations}
\begin{align}
\bvt^i        &= e^i_\mu\,dx^\mu\,,\label{E:four coframes}\\
\bvt^{\hfive} &= e^{\hfive}_\mu\,dx^\mu + e^{\hfive}_5\,dx^5
               = -\kappa\,\phi\,A_{\mu}\,dx^\mu + \phi\,dy \label{E:fifth coframe}
\end{align}
\end{subequations}
with $y\equiv x^5$.
From (\ref{E:four coframes}) and (\ref{E:fifth coframe}), 
we have the inverse relation
\begin{equation}
dy  = e^5_i\,\bvt^i + e^5_{\hfive}\,\bvt^{\hfive}
    = \kappa\,A_{\mu}\,e^{\mu}_i\,\bvt^i + \frac{1}{\phi}\,\bvt^{\hfive}\,.
\end{equation}
The components of the torsion tensor 
in the orthonormal frame are obtained by
\begingroup
\allowdisplaybreaks
\begin{subequations}\label{E:torsion components}
\begin{align}
\bar{T}^{i}{}_{jk}           &= T^{i}{}_{jk} + 
                                \kappa\,A_{\mu}(\partial_{5}e^{i}_{\nu})
                                (e^{\mu}_{j}e^{\nu}_{k}-e^{\mu}_{k}e^{\nu}_{j})\,,\label{E:induced torsion}\\
\bar{T}^{i}{}_{\hfive j}     &= \frac{1}{\phi}(\partial_{5}e^{i}_{\mu})e^{\mu}_{j}\,,\\
\bar{T}^{\hfive}{}_{ij}      &= -\frac{\kappa}{2}\,\phi\,e^{\mu}_{i}e^{\nu}_{j}\,F_{\mu\nu}+
                                \kappa^2\,\phi\,A_{\mu}(\partial_{5}\,A_{\nu})
                                (e^{\nu}_{i}e^{\mu}_{j}-e^{\nu}_{j}e^{\mu}_{i}),\\
\bar{T}^{\hfive}{}_{i\hfive} &= \frac{1}{\phi}e^{\mu}_{i}(\partial_{\mu}\phi) +
                                \frac{1}{\phi}\kappa\,A_{\mu}\,e^{\mu}_{i}(\partial_{5}\phi) +
                                \kappa\,e^{\mu}_{i}(\partial_{5}A_{\mu})\,,
\end{align}
\end{subequations}
\endgroup
where $F_{\mu\nu}:=\partial_{\mu}A_{\nu}-\partial_{\nu}A_{\mu}$
and $T^{i}{}_{jk} = e^{\mu}_{j}(\partial_{\mu}e^{i}_{\nu})e^{\nu}_{k}
- e^{\mu}_{k}(\partial_{\mu}e^{i}_{\nu})e^{\nu}_{j}$, 
which are four-dimensional quantities without involving the fifth component. 

The \emph{extended} torsion scalar in $D$-dimension, 
${}^{(D)}T^{(\text{ext})}$, 
can be shown in a similar way as (\ref{E:NGR Lagrangian}).
In particular, the five-dimensional extended torsion scalar is given by
\begin{equation}\label{E:5d ext torsion scalar in o-frame}
{}^{(5)}T^{(\text{ext})} = a\,\bar{T}_{LMN}\bar{T}^{LMN} + 
            b\,\bar{T}_{LMN}\bar{T}^{NML} + 
            c\,\bar{T}^{L}{}_{LM}\bar{T}^{N}{}_{N}{}^{M}\,.
\end{equation}
We decompose the contraction of the fifth component part 
from the extended torsion scalar, expressed as
\begin{align}\label{E:5d dec-ext torsion scalar in o-frame}
{}^{(5)}T^{(\text{ext})} 
=& \,\bar{T}_{\text{NGR}} 
   + 2a\,\bar{T}_{i\hfive j} \bar{T}^{i\hfive j} 
   + a\,\bar{T}_{\hfive ij} \bar{T}^{\hfive ij} 
   + b\,\bar{T}_{i\hfive j} \bar{T}^{j\hfive i} 
   + 2b\,\bar{T}_{\hfive ij} \bar{T}^{ji\hfive} \nonumber\\
 & +(2a+b+c)\,\bar{T}_{\hfive i\hfive} \bar{T}^{\hfive i\hfive} 
   + 2c\,\bar{T}^{j}{}_{j}{}^{i} \bar{T}^{\hfive}{}_{\hfive}{}^{i} 
   +c\,\bar{T}^{i}{}_{i\hfive} \bar{T}^{j}{}_{j}{}^{\hfive}\,,
\end{align}
where the torsion scalar $\bar{T}_{\text{NGR}}$
is in the five-dimensional geometric form, 
constructed by $\bar{T}^{i}{}_{jk}$ in (\ref{E:induced torsion}).
According to (\ref{E:torsion components}), we can write down 
8 terms for each extended five-dimensional torsion scalar 
(\ref{E:5d dec-ext torsion scalar in o-frame}),
as shown in Appendix~\ref{A:decomposition terms}.

One can consider the $n$-mode \emph{harmonic expansions} 
with the fifth dimensional radius $r$ for  
$e^{i}_{\mu}$, $A_{\mu}$ and $\phi$, given by
\begin{subequations}\label{E:harmonic expansions}
\begin{align}
e^{i}_{\mu}(x^{\mu},y) &= \sum_n e^{i(n)}_{\mu}(x^{\mu})\,e^{iny/2r}\,, \\ 
A_{\mu}(x^{\mu},y)     &= \sum_n A_{\mu}^{(n)}(x^{\mu})\,e^{iny/r}\,, \\
\phi(x^{\mu},y)        &= \sum_n \phi^{(n)}(x^{\mu})\,e^{iny/r}\,,
\end{align}
\end{subequations}
respectively.
We specify the physical gravitational fields as 
the metric tensor with the harmonic expansion of the form
$g_{\mu\nu}(x,y)=\sum_n g^{(n)}_{\mu\nu}(x)\,e^{iny/r}$.
Similarly, the harmonic expansion of $e^{\mu}_{i}$ is defined by
$e^{\mu}_{i}(x,y) = \sum_n e^{\mu(n)}_{i}(x)\,e^{-iny/2r}$ 
in order to satisfy the orthonormality relation.
The mass squared $n^2/r^2$ will be generated by 
the derivative respect to the $y$ coordinate.

The action of the zeroth KK mode ($n=0$) 
in the four-dimensional effective extended gravity
in teleparallism can be written as
\begin{align}\label{E:zero mode action}
S_{g}^{(0)} 
=& \int d^4x\, e^{(0)}
  \bigg(\frac{1}{2\kappa_4}\phi^{(0)}\, T^{(0)}_{\text{NGR}} 
  - \frac{a\kappa^2}{8\kappa_4}\phi^{(0)\,3}\,g^{(0)\,\mu\rho}g^{(0)\,\nu\sigma}
  F^{(0)}_{\mu\nu}F^{(0)}_{\rho\sigma} \nonumber\\
 &+ \frac{2a+b+c}{2\kappa_4}\frac{1}{\phi^{(0)}}\,g^{(0)\,\mu\nu}
  \partial_{\mu}\phi^{(0)}\partial_{\nu}\phi^{(0)}
  - \frac{c}{\kappa_4}\,g^{(0)\,\mu\nu}\,T^{(0)}_{\mu}\,\partial_{\nu}\phi^{(0)}\bigg)\,,
\end{align}
where $\kappa_4:= \k5/\lambda= 8\pi\,G_5/\lambda$
is the effective gravitational constant
with the fifth dimensional integral factor
$\lambda\equiv2\pi r$, 
$\kappa$ can be chosen as $\sqrt{\kappa_4}$,
and $T^{(0)}_{\mu}:=T^{(0)\,\nu}{}_{\nu\mu}$ is the torsion vector.
The effective action (\ref{E:zero mode action}) is invariant under 
the global conformal transformation in the absence of the 
vector field $A^{(0)}_{\mu}$, which can be identified as the electromagnetic field, 
while the associated Noether current density is 
$J^{(0)}_{\mu}:=e(2a+b+c)\partial_{\mu}\phi^{(0)}/\kappa_4$.

Note that in the case of the five-dimensional TEGR theory,
the action (\ref{E:zero mode action}) with $(a,b,c)=(1/4,1/2,-1)$
can be reduced to those in \cite{deAndrade:1999vq} 
and \cite{Geng:2014yya, Geng:2014nfa} with $\phi^{(0)} =1$ and $A_{\mu}^{(0)} =0$, respectively.

\textbf{(ii) Matter part}

We now add the matter Lagrangian 
${}^{(5)}\lagrangian_{\text{m}}={}^{(5)}e\,L_{\text{m}}(e^{I}_{M},\Psi,\mathcal{D}_{M}\Psi)$ 
into the effective action (\ref{E:zero mode action}),
where $\mathcal{D}_{M}$ is the covariant derivative of the matter field $\Psi$.
The $n$-mode harmonic expansion of $\Psi$ is given by
\begin{equation}
\Psi(x^{\mu},y) = \sum_n \Psi(x^{\mu})\,e^{iny/r}\,.
\end{equation}
The term $\mathcal{D}_{5}\Psi$ will produce the KK mass terms for the $\Psi$ field.
For the massless zero mode, the matter field $\Psi^{(0)}$ 
is assumed to be localized on the $T_4$ hypersurface.
After the KK reduction, the effective matter Lagrangian becomes
\begin{equation}
\mathcal{L}^{(0)}_{\text{m,\,eff}}
= e^{(0)}\lambda\phi^{(0)}L_{\text{m}}({e^{i(0)}_{\mu}},A^{(0)}_{\mu},
\phi^{(0)},\Psi^{(0)},\mathcal{D}_{\mu}\Psi^{(0)})\,.
\end{equation}

Without confusion, we shall drop all the superscript $(0)$ 
in our discussion for the four-dimensional effective theory.
We can vary the full action $S=S_{g}+S_{\text{m}}$ with respect to 
$e^{i}_{\mu}$, $A_{\mu}$ and $\phi$, resulting in
\begingroup
\allowdisplaybreaks
\begin{subequations}
\begin{align}
\frac{1}{2}\,e^{\mu}_{i}\bigg(
\phi\, T_{\text{NGR}} 
  - \frac{a\kappa^2}{4}\phi^{3}\,g^{\lambda\rho}g^{\nu\sigma}
  F_{\lambda\nu}\,F_{\rho\sigma} 
  + \frac{2a+b+c}{\phi}\,g^{\lambda\nu}
  \partial_{\lambda}\phi\,\partial_{\nu}\phi 
& \nonumber \\ 
  - 2\,c\,g^{\lambda\nu}\,T_{\lambda}\,\partial_{\nu}\phi\bigg)
  - e^{\rho}_{i}\bigg\{\phi\,T^{j}{}_{\rho\nu}\,\Sigma_{j}{}^{\mu\nu}
  - \frac{a\kappa^2}{2}\phi^{3}\,g^{\mu\lambda}g^{\nu\sigma}
    \,F_{\lambda\nu}\,F_{\rho\sigma}
& \nonumber \\ 
  + \frac{2a+b+c}{\phi}\,g^{\mu\lambda}
  \partial_{\lambda}\phi\,\partial_{\rho}\phi  
  - c\,\bigg(\partial_{\s}\phi\,T^{\mu}{}_{\rho}{}^{\s} 
  + \partial_{\rho}\phi\,T^{\mu} + \partial^{\mu}\phi\,T_{\rho}\bigg)\bigg\}
& \nonumber \\ 
  + \,\frac{1}{e}\, \partial_{\nu}\bigg\{e\,\bigg(\phi\,\Sigma_{i}{}^{\mu\nu} 
  - c\,e^{\mu}_{i}\,\partial^{\nu}\phi 
  + c\,e^{\nu}_{i}\,\partial^{\mu}\phi\bigg)\bigg\}
& = \kappa_4\lambda\phi\,\Theta^{\mu}_{i}\,, \\ 
  {\frac{\lambda}{e}\frac{\delta \lagrangian_{\text{m}}}{\delta A_{\nu}}}
    + \frac{1}{e}\frac{a\kappa^2}{2\kappa_4}\,
      \phi^2\,\partial_{\mu}\bigg(eF^{\mu\nu}\bigg)
    + \frac{3a\kappa^2}{2\kappa_4}\,\phi\,\partial_{\mu}\phi\,F^{\mu\nu} 
& = 0\,, \\
  T_{\text{NGR}}
  - \frac{3a\kappa^2}{4}\,\phi^2\,
    g^{\mu\rho}g^{\nu\sigma}F_{\mu\nu}F_{\rho\sigma}
  + 2\kappa_4\lambda\,\bigg(L_{\text{m}}
    + {\frac{\phi}{e}\frac{\delta \lagrangian_{\text{m}}}{\delta \phi}}\bigg)
& \nonumber \\
  + \frac{2a+b+c}{\phi^2}\,
    \bigg(g^{\mu\nu}\partial_{\mu}\phi\partial_{\nu}\phi
    - 2\phi\,\hat\Box\phi\bigg)
  + \frac{2c}{e}\,\partial_{\nu}\bigg(eg^{\mu\nu}T_{\mu}\bigg)
& = 0\,,
\end{align}
\end{subequations}
\endgroup
respectively, where 
$\hat\Box:=(1/e)\partial_{\nu}(eg^{\mu\nu}\partial_{\mu})$ 
and 
$\Theta^{\mu}_{i} := 
(-1/e)(\delta\,\lagrangian_m/\delta\,e^{i}_{\mu})$ 
is the dynamical energy-momentum tensor 
with $\lagrangian_{\text{m}}:=eL_{\text{m}}$.

\section{Specific Models}

We now concentrate on the four-dimensional effective theory
in (\ref{E:zero mode action}). 
With the conformal transformations of the metric tensor 
and frame fields to be
$\widetilde{g}_{\mu\nu}=\Omega^2(x)\,g_{\mu\nu}$
and  $\widetilde{e}_{\mu}^{i}=\Omega\, e_{\mu}^{i}$,
respectively, the corresponding matter Lagrangian is given by
\begin{equation}\label{E:Conf trans of matter lagrangian_}
\lambda\phi\lagrangian_{\text{m}}(g^{\mu\nu},A_{\mu},\phi,\Psi,\mathcal{D}_{\mu}\Psi)
=\lambda\phi\,\Omega^{-4}\widetilde{e}L_{\text{m}}
(\Omega^{2}\widetilde{g}^{\mu\nu},\widetilde{A}_{\mu},\widetilde{\phi},
\widetilde{\Psi},\widetilde{\mathcal{D}}_{\mu}\widetilde{\Psi})\,,
\end{equation}
where $e=\Omega^{-4}\widetilde{e}$. 
In the 4-dimensional spacetime, we have the conformal transformations
\begin{subequations}\label{E:4D Conformal torsion}
\begin{eqnarray}
\widetilde{T}^{i}{}_{\mu\nu} & = & \Omega\,T^{i}{}_{\mu\nu} +
(\partial_{\mu}\Omega)\,e_{\nu}^{i} - e_{\mu}^{i}\,(\partial_{\nu}\Omega)\,,
\label{E:4D Conformal torsion tensor}\\
\widetilde{T}_{\mu} & = & T_{\mu} - 3\,\Omega^{-1}\,\partial_{\mu}\Omega\,,
\label{E:4D Conformal torsion vector}
\end{eqnarray} 
\end{subequations}
for the torsion tensor and vector, respectively.
By extending the transformations to the torsion scalar of NGR, 
we obtain
\begin{align}\label{E:4D Conformal extended torsion scalar}
T_{\text{NGR}} =\,
& \Omega^2\,\widetilde{T}_{\text{NGR}} 
  + \bigg(4\,a + 2\,b + 6\,c\bigg)
     \widetilde{g}^{\mu\nu}\widetilde{T}_{\mu}(\Omega\,\partial_{\nu}\Omega) \nonumber\\
& + \bigg(6\,a + 3\,b + 9\,c\bigg)
     \widetilde{g}^{\mu\nu}\partial_{\mu}\Omega\,\partial_{\nu}\Omega \,.
\end{align}
In Appendix~\ref{B:D-dimensional conformal transformation}, 
we  show the formulas of the $D$-dimensional conformal transformation 
for the torsion tensor.

For the massless scalar field $\phi$, 
the conformal transformation is
$\widetilde{\phi}=\Omega^{\lambda_{\phi}}\,\phi$ 
with the conformal weight parameter $\lambda_{\phi}$.
On the other hand, one has
$\widetilde{A}_{\mu}=A_{\mu}$ and $\widetilde{F}_{\mu\nu}=F_{\mu\nu}$
for the vector field and field strength, 
which are independent of the metric, respectively.
Finally, the conformal transformed effective Lagrangian is found to be
\begingroup
\allowdisplaybreaks
\begin{align}\label{E:Conformal effective Lagrangian}
\mathcal{L}_{g} =\,\widetilde{e}\,\frac{\Omega^{-(\lambda_{\phi}+2)}}{2\kappa_4}\,
\bigg\{ 
&   \widetilde{\phi}\, \widetilde{T}_{\text{NGR}} 
  + \Omega^{-1}\,\bigg(4\,a+2\,b+(6+2\lambda_{\phi})\,c\bigg)\,
    \widetilde{\phi}\,\widetilde{g}^{\mu\nu}\,\widetilde{T}_{\mu}\,\partial_{\nu}\Omega
    \nonumber\\
& + \Omega^{-2}\,\bigg((6+2\lambda_{\phi}^2)\,a
        +(3+\lambda_{\phi}^2)\,b+(9+6\,\lambda_{\phi}+\lambda_{\phi}^2)\,c\bigg)
    \widetilde{\phi}\,\widetilde{g}^{\mu\nu}\,\partial_{\mu}\Omega\,\partial_{\nu}\Omega
    \nonumber\\
& - \Omega^{-2(\lambda_{\phi}-1)}\,\frac{a\kappa^2}{4}
    \widetilde{\phi}^{\,3}\,\widetilde{g}^{\mu\rho}\widetilde{g}^{\nu\sigma}
    \widetilde{F}_{\mu\nu}\widetilde{F}_{\rho\sigma} 
  + \bigg(2a+b+c\bigg)
    \frac{1}{\widetilde{\phi}}\,\widetilde{g}^{\mu\nu}
    \partial_{\mu}\widetilde{\phi}\,\partial_{\nu}\widetilde{\phi}
    \nonumber\\
& - 2\,c\,\widetilde{g}^{\mu\nu}\,\widetilde{T}_{\mu}\,\partial_{\nu}\widetilde{\phi}
  + \Omega^{-1}\,\bigg(-4\lambda_{\phi}\,a
        -2\,\lambda_{\phi}\,b-(2\,\lambda_{\phi}+6)\,c\bigg)\,
    \widetilde{g}^{\mu\nu}\,\partial_{\mu}\Omega\,\partial_{\nu}\widetilde{\phi}\bigg\}\,.
\end{align}
\endgroup

Due to the undetermined coefficients $(a,b,c)$,
we would examine the necessary conditions for having 
the Einstein-frame as well as conformal invariant models.
By taking $\lambda_{\phi}=-2$ with the conformal scalar $\omega:=\ln\Omega$,
the Lagrangian (\ref{E:Conformal effective Lagrangian}) becomes
\begingroup
\allowdisplaybreaks
\begin{align}\label{E:lambda=2 Conformal effective Lagrangian}
\mathcal{L}_{g} =\,\widetilde{e}\frac{1}{2\kappa_4}\,
\bigg\{ 
&   \widetilde{\phi}\, \widetilde{T}_{\text{NGR}} 
  + \bigg(4\,a+2\,b+2\,c\bigg)\,
    \widetilde{\phi}\,\widetilde{g}^{\mu\nu}\,\widetilde{T}_{\mu}\,\partial_{\nu}\omega
    \nonumber\\
& + \bigg(14\,a+7\,b+c\bigg)
    \widetilde{\phi}\,\widetilde{g}^{\mu\nu}\,\partial_{\mu}\omega\,\partial_{\nu}\omega
    \nonumber\\
& - e^{6\omega}\,\frac{a\kappa^2}{4}
    \widetilde{\phi}^{\,3}\,\widetilde{g}^{\mu\rho}\widetilde{g}^{\nu\sigma}
    \widetilde{F}_{\mu\nu}\widetilde{F}_{\rho\sigma} 
  + \bigg(2a+b+c\bigg)
    \frac{1}{\widetilde{\phi}}\,\widetilde{g}^{\mu\nu}
    \partial_{\mu}\widetilde{\phi}\,\partial_{\nu}\widetilde{\phi}
    \nonumber\\
& - 2\,c\,
    \widetilde{g}^{\mu\nu}\,\widetilde{T}_{\mu}\,\partial_{\nu}\widetilde{\phi}
  + \bigg(8\,a+4\,b-2\,c\bigg)\,
    \widetilde{g}^{\mu\nu}\,\partial_{\mu}\omega\,\partial_{\nu}\widetilde{\phi}\bigg\}\,.
\end{align}
\endgroup

\subsection{Einstein-Frame Condition}
In general, the Einstein-frame does not exist for 
the non-minimal torsion scalar $\phi T_{\text{NGR}}$
due to the non-minimal coupling $\widetilde{g}^{\mu\nu}\widetilde{T}_{\mu}(\Omega\,\partial_{\nu}\Omega)$ 
generated in (\ref{E:4D Conformal extended torsion scalar}),
which is different from the transformation of the curvature scalar in GR.
In order to transform the effective action (\ref{E:zero mode action})
from the Jordan-frame to Einstein-frame,
we only focus on the gravitational part and 
choose $\widetilde{\phi}=\Omega^{-2}\,\phi=1$.
If the coefficients in (\ref{E:lambda=2 Conformal effective Lagrangian}) 
satisfy the condition 
\begin{equation}\label{E:E-frame condition}
2\,a+b+c= 0\,,
\end{equation}
the Einstein-frame is obtained by eliminating the term
$\widetilde{g}^{\mu\nu}\widetilde{T}_{\mu}\partial_{\nu}\omega$.
As a result, the Lagrangian in (\ref{E:zero mode action}) is reduced to
\begin{equation}\label{E:Jordan of T_ngr}
\mathcal{L}_{g}
= e\,\frac{1}{2\kappa_4}  
  \bigg(\phi\,T_{\text{NGR}}
- 2\,c\,g^{\mu\nu}T_{\mu}\partial_{\nu}\phi
- \frac{a\kappa^2}{4}\,\phi^{3}\,
  g^{\mu\rho}g^{\nu\sigma}F_{\mu\nu}F_{\rho\sigma} \bigg)\,.
\end{equation}
In the corresponding Einstein-frame, it reads as
\begin{equation}\label{E:Einstein of T_ngr}
\mathcal{L}_{g}^{(\text{E})} 
= \widetilde{e}\,
  \bigg(\frac{1}{2\kappa_4}\widetilde{T}_{\text{NGR}}
- \frac{c}{2}\,\widetilde{g}^{\mu\nu}\partial_{\mu}\vphi\,\partial_{\nu}\vphi
- \frac{a\kappa^2}{8\kappa_4}\,e^{6\omega}\,
  \widetilde{g}^{\mu\rho}\widetilde{g}^{\nu\sigma}
  \widetilde{F}_{\mu\nu}\widetilde{F}_{\rho\sigma} \bigg)\,,
\end{equation}
where $\vphi:=\sqrt{6/\kappa_4}\,\omega$,
As seen from (\ref{E:Einstein of T_ngr}), 
to avoid the ghost field, 
one has an additional condition $c\leq0$.
As a result,
 the conditions 
\begin{equation}\label{E:Einstein-frame sol}
2a+b+c=0 \qquad \text{and} \qquad c\leq0
\end{equation} 
are needed to have  the Einstein-frame.
For a simple choice of $c=-1$, 
one gets the \emph{minimal coupled} one-parameter family model with $2a+b=1$
in teleparallelism~\cite{Hayashi:1979qx,Hehl:1978yt,Nitsch:1979qn}.

We note that the coefficients $(a,b,c)=(1/4,1/2,-1)$ satisfy
(\ref{E:E-frame condition}), which reduce 
(\ref{E:Jordan of T_ngr}) to be the case of TEGR in the KK theory with the gauge field $A_{\mu}$,
whereas that for the Lagrangian (\ref{E:Einstein of T_ngr}) of TEGR
in the Einstein-frame 
 without $A_{\mu}$  has been presented
in~\cite{Geng:2014nfa}.

\subsection{Conformal Invariant}\label{Subsec:Conformal Invariant}
It is apparent that (\ref{E:Conformal effective Lagrangian})
or (\ref{E:lambda=2 Conformal effective Lagrangian})
is not conformal invariant in the presence of the vector field $A_{\mu}$.
For the case without $A_{\mu}$, 
from (\ref{E:lambda=2 Conformal effective Lagrangian}) 
we obtain the requirements of a conformal invariant theory
by only keeping the terms of 
$\widetilde{\phi}^{-1}\,\widetilde{g}^{\mu\nu}
\partial_{\mu}\widetilde{\phi}\,\partial_{\nu}\widetilde{\phi}$ 
and 
$\widetilde{g}^{\mu\nu}\,\widetilde{T}_{\mu}\,\partial_{\nu}\widetilde{\phi}$, 
given by
\begin{subequations}\label{E:Conformal condition}
\begin{align}
4a   +2b  +2c   &= 0\,, \\
14a  +7b  +c    &= 0\,, \\
8a   +4b  -2c   &= 0\,.
\end{align}
\end{subequations}
The solution for (\ref{E:Conformal condition}) is 
\begin{equation}\label{E:Conformal sol}
2a+b=0 \qquad \text{and} \qquad c=0
\end{equation}
corresponding to the simple one-parameter 
conformal invariant gravity in teleparallelism.
The action is given by
\begin{align}\label{E:Conformal inv action}
S_{g}^{\text{(c)}} 
=& \int d^4x\,\bigg\{ e
  \frac{a}{2\kappa_4}\phi\,
    \bigg(T_{ijk} \, T^{ijk} 
    -2\,T_{ijk} \,T^{kji} \bigg)
    +\lambda\phi\lagrangian_{\text{m}}\bigg\}\,,
\end{align}
where $\phi$ is the auxiliary field, which can be identified as
the varying Newton's constant.
For the variation with respect to the frame fields $e^{i}_{\mu}$, 
we get the gravitational equation of motion.
\begin{align}\label{E:Conformal gravi eom}
& \phi\,\bigg\{
  \frac{1}{2}\,e^{\mu}_{i}\,T^{j}{}_{\rho\nu}\,
  \bigg(T_{j}{}^{\rho\nu} - 2\,T^{\nu\rho}{}_{j}\bigg)
 - 4\,e^{\rho}_{i}\,T^{j}{}_{\rho\nu}\,K^{\mu\nu}{}_{j}\bigg\}\nonumber\\
&+ \frac{2\phi}{e}\,\partial_{\nu}\bigg\{
   e\,\bigg(T_{i}{}^{\mu\nu} - 2\,T^{\nu\mu}{}_{i}\bigg)\bigg\}
 + 2\,\bigg(T_{i}{}^{\mu\nu} - 2\,T^{\nu\mu}{}_{i}\bigg)\partial_{\nu}\phi
 = \kappa_4\lambda\phi\,\Theta^{\mu}_{i}\,.
\end{align}
Furthermore, the equation of motion of $\phi$ leads to the constraint 
\begin{equation}\label{E:Conformal gravity constraint}
{a\,T_{ijk}(T^{ijk} - 2\,T^{kji}) 
+ 2\kappa_4 \lambda\bigg(L_{\text{m}}
+ \frac{\phi}{e}\frac{\delta \lagrangian_{\text{m}}}{\delta \phi}\bigg) = 0\,.}
\end{equation}
In a pure gravity theory, 
(\ref{E:Conformal gravity constraint}) yields
(i) $T_{ijk}=0$ or
(ii) $T^{ijk} = 2\,T^{kji}$.
Note that  (i) is the torsion-free case, which is forbidden due to no gravity.
For (ii), it implies that $T^{iik}=2T^{kii}=0$,
which contains no torsion vector mode, 
so that the equation of motion (\ref{E:Conformal gravi eom}) reads 
\begin{equation}
e^{\rho}_{i}\,T^{j}{}_{\rho\nu}\,T^{\nu\mu}{}_{j}=0\,.
\end{equation}

On the other hand, if we assume that $\omega=\omega(\phi)$,
under the conformal transformation, 
$\widetilde{\phi}$ could be expressed as the function $\widetilde{\phi}(\phi)$,
as long as $\det(d\widetilde{\phi}/d\phi)$ exists. Subsequently, 
we can have $\phi(\widetilde{\phi})$ by the inverse function theorem,
resulting in $\omega(\widetilde{\phi})$ by 
$d\widetilde{\phi}/d\phi=\omega' \exp(\lambda_{\phi}\omega)(1+\lambda_{\phi}\omega'\phi)$
with $\omega':=d\omega/d\phi$.
By setting 
\begin{equation}
\partial_{\mu}\omega = \partial_{\mu}\ln\widetilde{\phi}\,,
\end{equation}
the Lagrangian density (\ref{E:lambda=2 Conformal effective Lagrangian}) 
without $A_{\mu}$ reads
\begin{align}\label{E:Conformal inv Lagrangian 2}
\mathcal{L}_{g} =\,\widetilde{e}\frac{1}{2\kappa_4}\,
\bigg\{ 
   \widetilde{\phi}\, \widetilde{T}_{\text{NGR}} 
  + \bigg(4\,a+2\,b\bigg)\,
    \widetilde{g}^{\mu\nu}\,\widetilde{T}_{\mu}\,\partial_{\nu}\widetilde{\phi}
 + \bigg(24a+12b\bigg)
    \frac{1}{\widetilde{\phi}}\,\widetilde{g}^{\mu\nu}
    \partial_{\mu}\widetilde{\phi}\,\partial_{\nu}\widetilde{\phi}\bigg\}\,.
\end{align}
By comparing the coefficients in (\ref{E:zero mode action}), 
the new requirements of the conformal invariance are found to be
\begin{subequations}\label{E:Conformal condition 2}
\begin{align}
2a    +b     &=  -c          \,, \\
24a   +12b   &=  2a   +b  +c \,,
\end{align}
\end{subequations}
leading to
the solution of $2a+b=0$ and $c=0$ as (\ref{E:Conformal sol}). 
It is apparent that the conformal invariance gives rise to  $2a+b+c=0$, which is the same as 
(\ref{E:E-frame condition}) for
the existence of the Einstein-frame.

\subsection{Weyl Gauge Invariant}
In this subsection, we would like to discuss the \emph{gauge theory}
in which the \emph{gauge field} would be introduced 
from the covariant derivative of the scalar filed $\phi$.
According to (\ref{E:zero mode action}) without the $A_{\mu}$ field, 
the Lagrangian can be rewritten as
\begin{align}\label{E:Weyl inv Lagrangian}
\mathcal{L}_{g} = \,e\frac{1}{2\kappa_4}
 & \bigg\{
   \phi\, \bigg(T_{\text{NGR}} 
   - kc\,g^{\mu\nu}\,T_{\mu}T_{\nu}\bigg)
\nonumber\\
 & + \frac{c}{k\phi}\,\bigg(g^{\mu\nu}
   (\partial_{\mu}-kT_{\mu})\phi
   (\partial_{\nu}-kT_{\nu})\phi\bigg)\bigg\}\,,
\end{align}
where 
$k=c/(2a+b+c)$ 
is a fixed ratio 
with $2a+b+c\neq0$ and $c\neq0$,
satisfying the equation
\begin{equation}\label{E:k condition}
\frac{1}{k}+2-3k=0\,,
\end{equation}
as shown in Appendix~\ref{A:CGIC}. 
We remark that $c/k=2a+b+c>0$ is the ghost-free condition for this model.
As a result, we obtain the conformal gauge invariant condition
to be either
\begin{equation}\label{E:Weyl gauge sol 1}
2a+b+4c =0
\end{equation}
or
\begin{equation}\label{E:Weyl gauge sol 2}
2a+b    =0
\end{equation}
with the constraints
\begin{equation}\label{E:Weyl gauge constraints}
2a+b+c>0 \qquad \text{and} \qquad c\neq0\,.
\end{equation}
The torsion vector $T_{\mu}$ in (\ref{E:Weyl inv Lagrangian})
can be identified as the gauge field under the gauge principle for the conformal/Weyl transformation.
Under the transformations
\begin{equation}
g_{\mu\nu} \longrightarrow e^{2\omega} g_{\mu\nu}\,,
\qquad
T_{\mu} \longrightarrow T_{\mu} - 3\,\partial_{\mu}\omega\,,
\qquad
\phi \longrightarrow e^{-2\omega}\,\phi\,,
\end{equation} 
the modified derivative of the $\psi$ field can be defined by
${}^{*}\partial^{(\psi)}_{\mu} 
= \partial_{\mu} 
+ (\lambda_{\psi}k/2)T_{\mu}$,
where the weight parameter $\lambda_{\psi}$ is defined by 
the transformation $\exp(\lambda_{\psi}\omega)$.
In addition, we have $\lambda_{\phi}=-2$ for $\phi$.
It can be shown that 
${}^{*}\partial^{(e)}_{\mu}e^{i}_{\nu}
= \partial_{\mu}e^{i}_{\nu} + (k/2)T_{\mu}e^{i}_{\nu}$
due to $\lambda_{g}=2\lambda_{e}=2$.
We can define a modified covariant derivative for $\psi$ given by
${}^{*}\nabla^{(\psi)} = {}^{*}d^{(\psi)} + {}^{*}\Gamma$
with the connection 
${}^{*}\Gamma^{\rho}_{\nu\mu} 
= e^{\rho}_{i}\,{}^{*}\partial^{(e)}_{\mu}e^{i}_{\nu}
= \Gamma^{\rho}_{\nu\mu}+ (k/2)\delta^{\rho}_{\nu}T_{\mu}$.
 Under this new covariant derivative,
the nonmetricity vanishes, \ie,
\begin{equation}
{}^{*}\nabla^{(g)}_{\mu}g_{\nu\rho}
= {}^{*}\partial^{(g)}_{\mu} g_{\nu\rho}
  - {}^{*}\Gamma^{\sigma}_{\nu\mu} g_{\sigma\rho}
  - {}^{*}\Gamma^{\sigma}_{\rho\mu} g_{\nu\sigma} 
= \nabla_{\mu}g_{\nu\rho}
=0\,.
\end{equation}
Furthermore, we have
\begin{subequations}
\begin{align}
{}^{*}T^{\rho}{}_{\mu\nu}
&= T^{\rho}{}_{\mu\nu} 
   + \frac{k}{2}(\delta^{\rho}_{\nu}T_{\mu} 
   - \delta^{\rho}_{\mu}T_{\nu})\,, \\
{}^{*}T_{\mu}
&= (1 - \frac{3}{2}k)\,T_{\mu}\,, \label{E:T to *T} 
\end{align}
\end{subequations}
and we can define
${}^{*}\widetilde{T}^{\rho}{}_{\mu\nu}
:=\widetilde{({}^{*}T)}{}^{\rho}{}_{\mu\nu}$, read as
\begin{subequations}
\begin{align}
{}^{*}\widetilde{T}^{\rho}{}_{\mu\nu}
&= \widetilde{T}^{\rho}{}_{\mu\nu} 
   + \frac{k}{2}\,(\delta^{\rho}_{\nu}\widetilde{T}_{\mu} 
   - \delta^{\rho}_{\mu}\widetilde{T}_{\nu})\,, \\
{}^{*}\widetilde{T}_{\mu}
&= (1 - \frac{3}{2}k)\,\widetilde{T}_{\mu}\,.
\end{align}
\end{subequations}
Note that 
$\widetilde{({}^{*}T)}{}^{\rho}{}_{\mu\nu}
\neq {}^{*}\left(\widetilde{T}\right){}^{\rho}{}_{\mu\nu}$.
We can rewrite the derivative as
${}^{*}\partial^{(\phi)}_{\mu}
:=\partial_{\mu} - \lambda_{\phi}k^2\, {}^{*}T_{\mu}$ 
in (\ref{E:Weyl inv Lagrangian}) from (\ref{E:k condition}) and (\ref{E:T to *T}),
which leads to the modified covariant derivative
${}^{*}\nabla^{(\phi)}_{\mu}\phi
={}^{*}\partial^{(\phi)}_{\mu}\phi$
with the new connection 
${}^{*}\Gamma^{\rho}_{\nu\mu}$.
Under the conformal transformation, 
the nonmetricity with respect to the modified connection is
\begin{equation}
{}^{*}\widetilde{\nabla}^{(g)}_{\mu}\widetilde{g}_{\nu\rho}
= e^{2\omega}({}^{*}\nabla^{(g)}_{\mu} g_{\nu\rho})
=0\,,
\end{equation}
which implies that ${}^{*}\widetilde{\nabla}^{(g)}_{\mu}$ is an invariant covariant derivative.
With the condition (\ref{E:Weyl gauge sol 1}) 
or (\ref{E:Weyl gauge sol 2}),
we have 
\begingroup
\allowdisplaybreaks
\begin{align}
& \,e\,\phi\bigg(T_{\text{NGR}} 
   - kc\,g^{\mu\nu}\,T_{\mu}T_{\nu}\bigg)
 \nonumber \\
=& \,e\,\phi\bigg({}^{*}T_{\text{NGR}} 
   - kc\,g^{\mu\nu}\,{}^{*}T_{\mu}{}^{*}T_{\nu}\bigg) \nonumber \\
=& \,\widetilde{e}\,\widetilde{\phi}\bigg({}^{*}\widetilde{T}_{\text{NGR}} 
   - kc\,\widetilde{g}^{\mu\nu}\,{}^{*}\widetilde{T}_{\mu}{}^{*}\widetilde{T}_{\nu}\bigg)\,,
\end{align}
\endgroup
where we have used the relation
${}^{*}\widetilde{T}^{\rho}{}_{\mu\nu}
={}^{*}T^{\rho}{}_{\mu\nu} 
- (1/2)(\delta^{\rho}_{\nu}\partial_{\mu}\omega 
- \delta^{\rho}_{\mu}\partial_{\nu}\omega)$.
Consequently, (\ref{E:Weyl inv Lagrangian}) becomes
\begin{align}\label{E:lambda=2 Weyl inv Lagrangian}
\mathcal{L}_{g} =\,\widetilde{e}\frac{1}{2\kappa_4}\,
\bigg\{\widetilde{\phi}
    \bigg({}^{*}\widetilde{T}_{\text{NGR}} 
   - kc\,g^{\mu\nu}\,{}^{*}\widetilde{T}_{\mu}{}^{*}\widetilde{T}_{\nu}\bigg)
  + \frac{c}{k\widetilde{\phi}}\,\widetilde{g}^{\mu\nu}
    {}^{*}\widetilde{\nabla}^{(\phi)}_{\mu}\widetilde{\phi}\,
    {}^{*}\widetilde{\nabla}^{(\phi)}_{\nu}\widetilde{\phi}\bigg\}\,,
\end{align}
where the covariant derivative for $\widetilde{\phi}$ is given by
\begin{equation}
{}^{*}\widetilde{\nabla}^{(\phi)}_{\mu}
= {}^{*}\widetilde{\partial}^{(\phi)}_{\mu} 
   + \bigg(\frac{3}{2}\,\lambda_{\phi}k^2-\lambda_{\phi}\bigg)\partial_{\mu}\omega\,,
\end{equation}
with
${}^{*}\widetilde{\partial}^{(\phi)}_{\mu}
=\partial_{\mu} - \lambda_{\phi}k^2\, {}^{*}\widetilde{T}_{\mu}$.
which has the same form as $\phi$ under the transformation,
given by 
${}^{*}\widetilde{\nabla}^{(\phi)}_{\mu}\widetilde{\phi} 
= e^{-2\omega}{}^{*}\nabla^{(\phi)}_{\mu}\phi$.
Clearly, (\ref{E:lambda=2 Weyl inv Lagrangian}) 
describes a conformal invariant theory.
This can be identified as the \emph{Weyl gauge theory}
on the \emph{Weyl-Cartan geometry} $Y_4$~\cite{Fulton:1962bu,Hayashi:1978xu,Blagojevic:2002du}.

\begin{table}[t]
\caption{The model parameters $a,b,$ and $c$ 
in the torsion scalar $T_{\text{NGR}}$ with 
minimal  and non-minimal coupled models.}
\begin{ruledtabular}
\begin{tabular}{l l c l}
Model                    &  Class          & {\tabincell{c}
                                                {Additional\\[-1.5ex]
                                                 condition\\[0.4ex]}}      & Reference \\
\hline
\multirow{9}{*}{\tabincell{c}
{Minimal coupled\\[-1ex]
$\frac{1}{2\kappa}T_{\text{NGR}}$}}       
                         &  \multirow{2}{*}
                             {\tabincell{l}
                             {$2a+b+c=0$,\\[-1.5ex]
                              $(a,b,c)=
                              (\frac{1}{4},
                               \frac{1}{2},
                               -1$)\\[0.4ex]}}
                                             &  -                 &  Einstein~\cite{Einstein:teleaction}\\
\cline{4-4}
                         &                   &  -                 &  Cho~\cite{Cho:1975dh}\\
\cline{2-4}
                         & \multirow{3}{*}
                             {\tabincell{l}
                             {$2a+b+c=0$,\\[-1.5ex]
                              $c=-1$\\[0.4ex]}}
                                             &  -                 &  Hehl \etal~\cite{Hehl:1978yt}\\
\cline{4-4}
                         &                   &  -                 &  Nitsch and Hehl~\cite{Nitsch:1979qn}\\
\cline{4-4}
                         &                   &  -                 &  Hayashi and Shirafuji~\cite{Hayashi:1979qx}\\
\cline{2-4}
                         &  \multirow{2}{*}
                             {\tabincell{l}
                             {$2a+b+c=0$,\\[-1.5ex]
                              $(a,b,c)=
                               (\frac{1}{2},0,-1$)\\[0.4ex]}}
                                             &  \multirow{2}{*}
                                                 {\tabincell{l}
                                                 {Static isotropic metric\\[-1.5ex]
                                                  in Scherrer~\cite{Scherrer:1955}\\[0.4ex]}}
                                                                  &  Hehl \etal~\cite{Hehl:1978yt}\\
\cline{4-4}
                         &                   &                    &  Nitsch and Hehl~\cite{Nitsch:1979qn}\\
\cline{2-4}
                         &  \tabincell{l}
                             {$2a+b+c=0$,\\[-1.5ex]
                              $(a,b,c)=
                              (\frac{1}{4},
                               \frac{1}{2},
                               -1$)\\[0.4ex]}
                                             &  \multirow{3}{*}
                                                  {\vspace{0.4ex}Einstein-frame} 
                                                                  &  Geng \etal~\cite{Geng:2014nfa}\\
\cline{2-2}\cline{4-4}
                         &  \tabincell{l}
                             {$2a+b+c=0$,\\[-1.5ex]
                              $c\leq0$\\[0.4ex]}
                                             &                    &  Sec.~III-A\\
\hline
\multirow{5}{*}
{\tabincell{c}{Non-minimal coupled\\[-1.5ex]
(conformal invarience)\\[-1ex]
$\frac{1}{2\kappa}\phi T_{\text{NGR}}$}} 
                         &  $2a+b+3c=0$      &  {\tabincell{l}
                                                 {$k'g^{\mu\nu}D_{\mu}\phi D_{\nu}\phi$,\\[-1.5ex]
                                                   where $D_{\mu}:=\partial_{\mu}-\frac{1}{3}T_{\mu}$\\[-1.5ex]
                                                   with arbitrary $k'$}}
                                                                  &  Maluf and Faria~\cite{Maluf:2011kf}\\
\cline{2-4}
                         &  {\tabincell{l}
                             {$2a+b+c=0$,\\[-1.5ex]
                              $c=0$\\[0.4ex]}}
                                             &  -                 &  Sec.~III-B\\
\cline{2-4}
                         &  $2a+b+4c=0$      &  \multirow{2}{*}
                                                 {\tabincell{l}
                                                  {$2a+b+c>0$,\\[-1.5ex]
                                                   $c\neq0$\\[0.4ex]}}
                                                                  &  \multirow{2}{*}{Sec.~III-C}\\
\cline{2-2}
                         &  $2a+b=0$         &                    &  \\
\end{tabular}
\end{ruledtabular}
\label{table:I}
\end{table}
We note that the ghost-free condition $2a+b+c=0$ is not considered here 
due to the vanished kinetic term of $\phi$.
It can be found that the conformal invariant theory (\ref{E:Conformal inv action})
is a special case of the Weyl gauge invariant model 
with (\ref{E:Weyl gauge sol 1}), (\ref{E:Weyl gauge sol 2}) 
and the ghost-free condition $2a+b+c=0$, implying that $2a+b+c=0$ and $c=0$.
However, this conformal invariant theory  can \emph{not} be identified as the gauge theory.
See also the discussion for the case of $2a+b+c=0$ in Appendix~\ref{A:CGIC}.

We remark that the conformal invariant model in telaparallelism 
with the condition $2a+b+3c=0$ has been investigated by Maluf and Faria~\cite{Maluf:2011kf}.
In their discussion, 
a new parameter $k'$ for the scalar kinetic term is introduced,
resulting in a four-parameters model.
In our discussion, the coefficient of the kinetic term of $\phi$ is $2a+b+c$ so that
the conformal invariant theory is totally determined by 
three parameters only.



\section{Weak Field Approximation}
If we define a canonical field $\Phi:=\sqrt{\phi/\kappa_4}$, 
the action (\ref{E:zero mode action}) with matter  can be written as 
\begingroup
\allowdisplaybreaks
\begin{align}\label{E:Canonical zero mode action}
S
=& \,S_{g} +S_{\text{m}} \nonumber\\
=& \int d^4x\, \bigg\{e
   \bigg(\frac{1}{2}\,\Phi^{2}\, T_{\text{NGR}} 
   - \frac{a\kappa^2\kappa_4^2}{8}\, \Phi^{6}\,g^{\mu\rho}g^{\nu\sigma}
   F_{\mu\nu}F_{\rho\sigma} \nonumber\\
 & + (4a+2b+2c)\,g^{\mu\nu}
   \partial_{\mu}\Phi \partial_{\nu}\Phi
   - 2c\,g^{\mu\nu}\,T_{\mu}\,\Phi \partial_{\nu}\Phi\bigg)
   + \kappa_4\lambda\Phi^2\mathcal{L}_{\text{m}}\bigg\}\,.
\end{align}
\endgroup
In the weak field approximation, 
the frame field is given by $e^{i}_{\mu} = \delta^{i}_{\mu} + h^{i}{}_{\mu}$ 
with $\left| h^{i}{}_{\mu} \right|\ll 1$, 
while the metric tensor $g_{\mu\nu}=\eta_{\mu\nu}+\gamma_{\mu\nu}$ 
can be read as the Minkowski background $\eta_{\mu\nu}$
with a small symmetric fluctuation $\gamma_{\mu\nu}:=2\,h_{(\mu\nu)}$.
The tensor $h_{\mu\nu}$ can be decomposed into
the symmetric $\gamma_{\mu\nu}$ and 
anti-symmetric $a_{\mu\nu}:=h_{[\mu\nu]}$ parts,
given by $h_{\mu\nu}=(1/2)\gamma_{\mu\nu}+a_{\mu\nu}$.
The reciprocals of the frame fields $e^{i}_{\mu}$ are 
$e^{\mu}_{i} = \delta^{\mu}_{i} + f_{i}{}^{\mu}$
with $f_{i}{}^{\mu}=-h_{i}{}^{\mu}$ 
due to the orthonormality of the frame fields.
Then, the torsion tensor and vector are given by
$T^{\rho}{}_{\mu\nu} = 
\delta^{\rho}_{i}(\partial_{\mu}h^{i}{}_{\nu} 
- \partial_{\nu}h^{i}{}_{\mu}) + \mathcal{O}(h^2_{\mu\nu})$
and
$T_{\nu} = \partial_{\mu}h^{\mu}{}_{\nu}-\partial_{\nu}h
+ \mathcal{O}(h^2_{\mu\nu})$ 
at $\mathcal{O}(h_{\mu\nu})$, respectively,
where
$h\equiv \delta^{\mu}_{i}\,h^{i}{}_{\mu}=h^{\mu}{}_{\mu}=(1/2)\gamma$ 
with $e =1 + h + \mathcal{O}(h^2_{\mu\nu})$.
We expand the geometric part in the Lagrangian (\ref{E:Canonical zero mode action}) 
and keep terms only in the lowest order:
\begin{align}\label{E:weak canonical zero mode Lagrangian}
\lagrangian_{g} 
\approx & \,\frac{1}{2}\,\Phi^{2}\, T_{\text{NGR}}
   - \frac{a\kappa^2\kappa_4^2}{8}\,
     \Phi^{6}\eta^{\mu\rho}\eta^{\nu\sigma}F_{\mu\nu}F_{\rho\sigma} \nonumber \\
  &+ (4a+2b+2c)\,
     \eta^{\mu\nu}\partial_{\mu}\Phi \partial_{\nu}\Phi
   - 2c\,\eta^{\mu\nu}T_{\mu}\,\Phi \partial_{\nu}\Phi\,,
\end{align}
where
\begingroup
\allowdisplaybreaks
\begin{align}\label{E:weak NGR Lagrangian}
T_{\text{NGR}} = \frac{1}{4}\bigg(
 & (2a
   + b)\partial_{\mu}\gamma_{\nu\rho}\,\partial^{\mu}\gamma^{\nu\rho}
   - (2a+b)\partial_{\mu}\gamma_{\nu\rho}\,
      \partial^{\rho}\gamma^{\mu\nu} \nonumber\\
 & + c\,\partial^{\rho}\gamma_{\rho\mu}\,\partial_{\sigma}\gamma^{\sigma\mu}
   - 2c\,\partial_{\mu}\gamma\,\partial_{\rho}\gamma^{\rho\mu}
   + c\,\partial_{\mu}\gamma\,\partial^{\mu}\gamma \bigg) \nonumber\\
 & + (2a+b)\partial_{\mu}\gamma_{\nu\rho}\,\partial^{\nu}a^{\mu\rho}
   + c\,\partial^{\rho}\gamma_{\rho\mu}\,\partial_{\sigma}a^{\sigma\mu} 
   - c\,\partial_{\mu}\gamma\,\partial_{\rho}a^{\rho\mu} \nonumber\\
 & + (2a-b)\partial_{\mu}a_{\nu\rho}\,\partial^{\mu}a^{\nu\rho} 
   + (2a-3b)\partial_{\mu}a_{\nu\rho}\,\partial^{\rho}a^{\mu\nu}
   + c\,\partial^{\rho}a_{\rho\mu}\,\partial_{\sigma}a^{\sigma\mu}\,,
\end{align}
\endgroup
which is the well-known \emph{Fierz-Pauli Lagrangian}~\cite{Fierz:1939ix}
with the conditions of 
$(a,b,c)=(1/4,1/2,-1)$ and $a_{\mu\nu}=0$.
The Noether current density for the new scalar $\Phi$ 
becomes $J_{\mu}=(4a+2b+2c)\,j_{\mu}$
with $j_{\mu}:=\Phi\partial_{\mu}\Phi$.

\begin{figure}
\includegraphics[width=.4\textwidth]{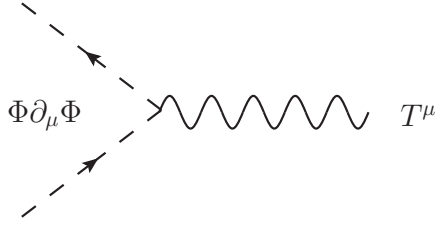}
\caption{
  The scalar current and torsion vector interaction 
  in the four-dimensional effective Lagrangian.} 
  \label{fig:I}
\end{figure}

Note that the last term of
$-2c\,T^{\mu}\Phi\partial_{\mu}\Phi$ in
(\ref{E:weak canonical zero mode Lagrangian})
gives the scalar current $j_{\mu}$ 
and torsion vector $T_{\mu}$ (current-vector) interaction
with the corresponding Feynman diagram shown in FIG.~\ref{fig:I}. 
By writing
$T_{\mu} \approx (1/2)\partial_{\rho}\gamma^{\rho}{}_{\mu} 
+ \partial_{\rho}a^{\rho}{}_{\mu}
- (1/2)\partial_{\mu}\gamma$,
this term contains the anti-symmetric tensor interaction 
$- 2c\,\partial_{\rho}a^{\rho\mu}\,\Phi \partial_{\mu}\Phi$,
reduced to
$c\,\Phi^2\,\partial_{\mu}\partial_{\rho}a^{\rho\mu}$
by using the integration by part,
which vanishes due to the symmetric property of the two derivatives.
As a result, there is no contribution 
in the current-vector interaction from $a_{\mu\nu}$.
However, the symmetric part cannot be eliminated 
even the \emph{de Donder gauge} 
$\partial_{\rho}\gamma^{\rho}{}_{\mu}-(1/2)\partial_{\mu}\gamma=0$ 
is imposed unless 
the transversed condition $\partial_{\rho}\gamma^{\rho}{}_{\mu}=0$ or
traceless condition $\gamma=0$ is used.
Clearly, it is different from GR in which the de Donder gauge would fix the degrees of freedom.
On the other hand, we can take the transverse gauges for
both symmetric and anti-symmetric parts, 
giving by $\partial_{\mu}\gamma^{\mu}{}_{\nu}=0$ and
$\partial_{\mu}a^{\mu}{}_{\nu}=0$, respectively.
By choosing 
$\gamma=0$, 
the torsion vector $T_{\mu}$ would vanish automatically,
resulting in that the degrees of freedom given by the frame fields in 
(\ref{E:weak canonical zero mode Lagrangian}) are the tensor modes only.

To simplify our calculation, we take that the variation of matter to be
$\delta\mathcal{L}_{\text{m}}
= (-1/2)\int T^{\mu\nu} \delta h_{\mu\nu}$ with
$T^{\mu\nu}:=-2\delta\mathcal{L}_{\text{m}}/\delta h_{\mu\nu}$.
Due to $h_{\mu\nu}=(1/2)\gamma_{\mu\nu}+a_{\mu\nu}$,
we define the energy-momentum tensors respect to $\gamma_{\mu\nu}$ and $a_{\mu\nu}$ are
$T^{\mu\nu}_{\gamma} := (1/2)T^{(\mu\nu)} 
= -2\delta\mathcal{L}_{\text{m}}/\delta\gamma_{\mu\nu}$ 
and 
$T^{\mu\nu}_{a} := T^{[\mu\nu]} 
= -2\delta\mathcal{L}_{\text{m}}/\delta a_{\mu\nu}$, respectively.
The corresponding equations of motion of 
(\ref{E:weak canonical zero mode Lagrangian})
for $\gamma_{\mu\nu}$, $a_{\mu\nu}$, $A_{\mu}$ and $\Phi$ are given by
\begingroup
\allowdisplaybreaks
\begin{subequations}
\begin{align}
  -j_{\rho}\bigg\{\frac{2a+b}{2}
     \partial^{\rho}\gamma^{\mu\nu} 
  -\frac{2a+b}{4}\bigg(\partial^{\nu}\gamma^{\rho\mu}
     +\partial^{\mu}\gamma^{\rho\nu}\bigg) 
& \nonumber \\
  +\frac{c}{4}\,\bigg(\eta^{\rho\mu}\partial_{\sigma}\gamma^{\sigma\nu}
     +\eta^{\rho\nu}\partial_{\sigma}\gamma^{\sigma\mu}\bigg) 
& \nonumber \\
  -\frac{c}{2}\,\bigg(\frac{1}{2}\eta^{\rho\mu}\partial^{\nu}\gamma
     +\frac{1}{2}\eta^{\rho\nu}\partial^{\mu}\gamma 
     +\eta^{\mu\nu}\partial_{\sigma}\gamma^{\sigma\rho}\bigg) 
& \nonumber \\
  +\frac{c}{2}\,\eta^{\mu\nu}\partial^{\rho}\gamma 
  +\frac{2a+b}{2}\bigg(\partial^{\mu}a^{\rho\nu}
     +\partial^{\nu}a^{\rho\mu}\bigg) & \nonumber \\
  +\frac{c}{2}\bigg(\eta^{\rho\mu}\partial_{\sigma}\a^{\sigma\nu}
     +\eta^{\rho\nu}\partial_{\sigma}\a^{\sigma\mu}\bigg)
  -c\eta^{\mu\nu}\partial_{\sigma}a^{\sigma\rho}\bigg\} 
& \nonumber \\
  -\Phi^2\bigg\{\frac{2a+b}{4}\,\Box\gamma^{\mu\nu}
  -\frac{2a+b}{8}
     \bigg(\partial_{\rho}\partial^{\nu}\gamma^{\mu\rho}
     +\partial_{\rho}\partial^{\mu}\gamma^{\nu\rho}\bigg)
& \nonumber \\
  +\frac{c}{8}
     \bigg(\partial^{\mu}\partial_{\sigma}\gamma^{\sigma\nu}
     +\partial^{\nu}\partial_{\sigma}\gamma^{\sigma\mu}\bigg) 
  -\frac{c}{4}\bigg(\partial^{\mu}\partial^{\nu}\gamma
     +\partial_{\rho}\partial_{\sigma}\gamma^{\sigma\rho}\eta^{\mu\nu}\bigg)
& \nonumber \\
  +\frac{c}{4}\,\eta^{\mu\nu}\Box\,\gamma 
  +\frac{2a+b}{4}\bigg(\partial_{\rho}\partial^{\mu}a^{\rho\nu}
     +\partial_{\rho}\partial^{\nu}a^{\rho\mu}\bigg) 
& \nonumber \\
  +\frac{c}{4}\bigg(\partial^{\mu}\partial_{\sigma}a^{\sigma\nu}
     +\partial^{\nu}\partial_{\sigma}a^{\sigma\mu}\bigg)\bigg\} 
  +\frac{c}{2}\,\partial^{\mu}j^{\nu} 
  +\frac{c}{2}\,\partial^{\nu}j^{\mu}
  -c\,\partial_{\rho}j^{\rho}\eta^{\mu\nu} 
& =\frac{1}{2}\kappa_4\lambda\Phi^2T^{\mu\nu}_{\gamma}\,, \label{E:Weak EoM of gamma}\\
  -j_{\rho}\bigg\{\frac{2a+b}{2}
     \bigg(\partial^{\mu}\gamma^{\rho\nu}-\partial^{\nu}\gamma^{\rho\mu}\bigg) 
  +\frac{c}{2}\bigg(\eta^{\rho\mu}\partial_{\sigma}\gamma^{\sigma\nu}
     -\eta^{\rho\nu}\partial_{\sigma}\gamma^{\sigma\mu}\bigg) 
& \nonumber \\
  -\frac{c}{2}\bigg(\eta^{\rho\mu}\partial^{\nu}\gamma
     -\eta^{\rho\nu}\partial^{\mu}\gamma\bigg)
  +2\bigg(2a-b\bigg)\partial^{\rho}a^{\mu\nu}
& \nonumber \\
  +\bigg(2a-3b\bigg)\bigg(\partial^{\mu}a^{\nu\rho}
     +\partial^{\nu}a^{\rho\mu}\bigg) 
  +c\,\bigg(\eta^{\rho\mu}\partial_{\sigma}a^{\sigma\nu}
     -\eta^{\rho\nu}\partial_{\sigma}a^{\sigma\mu}\bigg)\bigg\} 
& \nonumber \\
  -\Phi^2\bigg\{\frac{2a+b}{4}
     \bigg(\partial_{\rho}\partial^{\mu}\gamma^{\rho\nu}
     -\partial_{\rho}\partial^{\nu}\gamma^{\rho\mu}\bigg) 
  +\frac{c}{4}
     \bigg(\partial^{\mu}\partial_{\sigma}\gamma^{\sigma\nu}
     -\partial^{\nu}\partial_{\sigma}\gamma^{\sigma\mu}\bigg) 
& \nonumber \\
  +\bigg(2a-b\bigg)\Box\,a^{\mu\nu} 
  +\frac{2a-3b-c}{2}
     \bigg(\partial_{\rho}\partial^{\mu}a^{\nu\rho}
     +\partial_{\rho}\partial^{\nu}a^{\rho\mu}\bigg)\bigg\} 
& =\frac{1}{2}\kappa_4\lambda\Phi^2T^{\mu\nu}_{a}\,, \label{E:Weak EoM of a}\\
  \kappa_4\lambda\Phi^2{\frac{\delta \mathcal{L}_{\text{m}}}{\delta A_{\mu}}}
  + 3a\kappa^2\kappa_4^2\Phi^5(\partial_{\nu}\Phi)F^{\mu\nu}
  + \frac{a\kappa^2\kappa_4^2}{2}\Phi^6\partial_{\nu}F^{\mu\nu} 
&= 0\,, \label{E:Weak EoM of A}\\
  \Phi T_{\text{{NGR}}}
  - \frac{3a\kappa^2\kappa_4^2}{4}\Phi^5F_{\mu\nu}F^{\mu\nu}
  + 2\lambda\kappa_{4}\Phi\mathcal{L}_{\text{m}}
& \nonumber \\
  + \kappa_{4}\lambda\Phi^2{\frac{\delta \mathcal{L}_{\text{m}}}{\delta \Phi}}
  - \bigg(8a+4b+4c\bigg)\Box\,\Phi
  + 2c\,\Phi\partial_{\mu}T^{\mu} 
&= 0\,, \label{E:Weak EoM of phi}
\end{align}
\end{subequations}
\endgroup
respectively, where
$\Box:=\eta^{\mu\nu}\partial_{\mu}\partial_{\nu}$.

We now consider the conformal invariant model with 
the conditions 
in (\ref{E:Weyl gauge sol 1}) and (\ref{E:Weyl gauge sol 2}). 
From (\ref{E:Weak EoM of gamma}) and (\ref{E:Weak EoM of a}),
along with the gauge conditions 
$\partial_{\mu}\gamma^{\mu\nu}=0$, $\partial_{\mu}a^{\mu\nu}=0$ and $\gamma=0$,
we obtain 
\begingroup
\allowdisplaybreaks
\begin{subequations}\label{E:EoM of k=-1/3}
\begin{align}
  cj_{\rho}\bigg\{2\partial^{\rho}\gamma^{\mu\nu}
    -\partial^{\mu}\gamma^{\rho\nu}
    -\partial^{\nu}\gamma^{\rho\mu}
    +2\partial^{\mu}a^{\rho\nu}
    +2\partial^{\nu}a^{\rho\mu}\bigg\}
\nonumber\\
  +c\,\Phi^2\Box\,\gamma^{\mu\nu}
  +\frac{c}{2}\,\partial^{\mu}j^{\nu} 
  +\frac{c}{2}\,\partial^{\nu}j^{\mu}
  -c\,\partial_{\rho}j^{\rho}\eta^{\mu\nu} 
&=\frac{1}{2}\kappa_4\lambda\Phi^2T^{\mu\nu}_{\gamma}\,, \label{E:k=-1/3 weak conformal EoM of gamma} \\
  j_{\rho}\bigg\{2c\,\bigg(\partial^{\mu}\gamma^{\rho\nu}
  -\partial^{\nu}\gamma^{\rho\mu}\bigg)
  +4(b+2c)\partial^{\rho}a^{\mu\nu}
\nonumber\\
  +4\bigg(b+c\bigg)
  \bigg(\partial^{\mu}a^{\nu\rho}+\partial^{\nu}a^{\rho\mu}\bigg)\bigg\}
  +2\bigg(b+2c\bigg)\Phi^2\Box\,a^{\mu\nu}
&=\frac{1}{2}\kappa_4\lambda\Phi^2T^{\mu\nu}_{a} \,, \label{E:k=-1/3 weak conformal EoM of a}
\end{align}
\end{subequations}
\endgroup
for $2a+b+4c=0$ and  
\begin{subequations}\label{E:EoM of k=1}
\begin{align}
\frac{c}{2}\,\partial^{\mu}j^{\nu} 
  +\frac{c}{2}\,\partial^{\nu}j^{\mu}
  -c\,\partial_{\rho}j^{\rho}\eta^{\mu\nu} 
&=\frac{1}{2}\kappa_4\lambda\Phi^2T^{\mu\nu}_{\gamma}\,, \label{E:k=1 weak conformal EoM of gamma} \\
  -8aj_{\rho}f^{\rho\mu\nu}
  -4a\,\Phi^2\Box\,a^{\mu\nu}
&=\frac{1}{2}\kappa_4\lambda\Phi^2T^{\mu\nu}_{a} \,, \label{E:k=1 weak conformal EoM of a}
\end{align}
\end{subequations}
for $2a+b=0$, 
where the totally anti-symmetric tensor
$f^{\rho\mu\nu}:=\partial^{\rho}a^{\mu\nu}
+ \partial^{\mu}a^{\nu\rho}
+ \partial^{\nu}a^{\rho\mu}$ 
is the field strength of $a^{\mu\nu}$.
If we assume that the scalar field varies slowly, 
\ie, $\Phi\approx\Phi_c$ being a constant field, 
we have $j^{\mu}\approx0$, 
 reducing (\ref{E:EoM of k=-1/3}) and (\ref{E:EoM of k=1}) to be
\begin{subequations}\label{E:EoM of k=-1/3 with Phi_c}
\begin{align}
  \Box\,\gamma^{\mu\nu}
&=\frac{\kappa_4\lambda}{2c}\, T^{\mu\nu}_{\gamma}\,, \label{E:EoM of gamma k=-1/3 with Phi_c}\\
  \Box\,a^{\mu\nu}
&=\frac{\kappa_4\lambda}{4(b+2c)}\, T^{\mu\nu}_{a} \,, 
\end{align}
\end{subequations}
and 
\begin{subequations}\label{E:EoM of k=1 with Phi_c}
\begin{align}
0
&=T^{\mu\nu}_{\gamma}\,, \\
  \Box\,a^{\mu\nu}
&=-\frac{\kappa_4\lambda}{8a}\, T^{\mu\nu}_{a} \,,
\end{align}
\end{subequations}
respectively.
The Newtonian force can be identified in (\ref{E:EoM of gamma k=-1/3 with Phi_c}) by defining
$\kappa_{4,\text{eff}}:=8\pi\,G_{4,\text{eff}}=\kappa_4\lambda/c$
with  the effective Newtonian constant $G_{4,\text{eff}}=G_5/c$.
It is interesting to note that
only the anti-symmetric tensor $a^{\mu\nu}$ survives 
without the symmetric tensor field $\gamma^{\mu\nu}$
in 
(\ref{E:EoM of k=1 with Phi_c}) for the  $2a+b=0$ case,
which is different from that in TEGR.


\section{Conclusions}

We have presented the five-dimensional KK theory in teleparallelism 
including the vector $A_{\mu}$ and scalar $\phi$ fields. 
The model has been generalized in a similar way as 
NGR with the arbitrary coefficients $a, b$ and $c$.
We have decomposed the terms associated with 
the fifth dimension from the five-dimensional Lagrangian, 
as explicitly shown in Appendix~\ref{A:decomposition terms}.
We have concentrated on the zero mode in (\ref{E:zero mode action}) 
and found that the coupled term between the scalar derivative 
and torsion vector is generated after the KK reduction, 
which is different from the result in GR.

We have summarized 
the possible choices of the coefficients $(a,b,c)$ in TABLE~\ref{table:I} for various models in 
the literature.
To realize the Einstein-frame for the action in the four-dimensional effective action in  (\ref{E:zero mode action}), 
we have applied the conformal transformation and eliminated the 
gravitational non-minimal coupling terms by taking $2a+b+c=0$ with $c\leq0$.
By choosing $c=-1$, the Einstein-frame can be joined to the 
one-parameter teleparallel models~\cite{Hayashi:1979qx,Hehl:1978yt,Nitsch:1979qn}. 
For the conformal invariant model without the electromagnetic field, 
we have obtained the condition of  $2a+b=0$ with $c=0$,
corresponding to a pure gravity model.
Once the gauge field is introduced, 
we would construct  the conformal gauge theory.
Under the ghost-free constraints $2a+b+c>0$ and $c\neq0$ of (\ref{E:Weyl gauge constraints}), 
the non-minimal coupling terms of the torsion vector
can be absorbed in the  Kinetic term of the scalar field $\phi$ by using
the modified covariant derivative 
${}^{*}\nabla^{(\phi)}_{\mu}\phi:=(\partial_{\mu}+(\lambda_{\phi}k/2)T_{\mu})\phi$
with $\lambda_{\phi}=-2$ and
the torsion vector $T_{\mu}$ representing the Weyl gauge potential.
In order to satisfy the conformal gauge invariance, 
the new condition  of $2a+b+4c=0$ or $2a+b=0$ without $A_{\mu}$ has been found.
In such a modification, the geometry can be regarded as the Weyl-Cartan geometry $Y_4$, 
resulting in a Weyl gauge theory.
We have also shown that the ghost-free conditions for 
the theories with the Einstein-frame
and conformal invariant are 
$c\leq0$ and $2a+b+c>0$, respectively.

In terms of the redefinition of the scalar field 
$\Phi:=\sqrt{\phi/\kappa_4}$ 
with the mass dimension one, 
the non-minimal coupling of the torsion vector
can be identified as the interaction of the scalar current 
and torsion vector. 
For this interaction, we have considered 
the weak field limit $e^{i}_{\mu}=\delta^{i}_{\mu}+h^{i}{}_{\mu}$
with the gravitational waves 
$\gamma_{\mu\nu}=g_{\mu\nu}-\eta_{\mu\nu}=2\,h_{(\mu\nu)}$,
in which the tensor $h_{\mu\nu}=(1/2)\gamma_{\mu\nu}+a_{\mu\nu}$ 
contains both symmetric and anti-symmetric parts. 
We have shown that the anti-symmetric tensor $a_{\mu\nu}$
does not contribute to the current-vector interaction.
For the conformal invariant theory with $2a+b+4c=0$ and the slowly varying $\Phi$,
the effective Newtonian constant is independent of the fifth dimensional radius $r$
of $G_{4,\text{eff}}=G_5/c$.
Within the condition $2a+b=0$,
we have demonstrated that only the massless anti-symmetric tensor mode $a_{\mu\nu}$ 
can exist in the conformal invariant theory.


\begin{acknowledgments}
The work was partially supported by National Center for Theoretical Sciences 
and Ministry of Science and Technology (MOST 104-2112-M-009-020-MY3).
\end{acknowledgments}

\appendix
\begin{section}{The Decomposition Terms}\label{A:decomposition terms}
We show the decompositions of the extended five-dimensional torsion scalar as follows:
\begingroup
\allowdisplaybreaks
\begin{align}
\bar{T}_{\text{NGR}} 
=\,  & T_{\text{NGR}} 
       + 4a\kappa\,T_{l}{}^{\rho\sigma}A_{\rho}(\partial_{5}e^{l}_{\sigma})
       - 2a\kappa^2(g^{\mu\rho}A_{\mu}A_{\rho})
         (g^{\nu\sigma}\eta_{il}(\partial_{5}e^{i}_{\nu})
         (\partial_{5}e^{l}_{\sigma})) \nonumber\\
     & - 2a\kappa^2\eta_{il}(g^{\nu\rho}A_{\rho}(\partial_{5}e^{i}_{\nu}))
        (g^{\mu\sigma}A_{\mu}(\partial_{5}e^{l}_{\sigma}))
       + 2b\kappa T^{\sigma\rho}{}_{k}A_{\rho}(\partial_{5}e^{k}_{\sigma}) \nonumber\\
     & - 2b\kappa T^{\rho\sigma}{}_{k}A_{\rho}(\partial_{5}e^{k}_{\sigma})
       + b\kappa^2(g^{\mu\rho}A_{\mu}A_{\rho})(\partial_{5}e^{i}_{\nu})
         (\partial_{5}e^{k}_{\sigma})e^{\nu}_{k}e^{\sigma}_{i} \nonumber\\
     & - 2b\kappa^2(g^{\mu\sigma}A_{\mu}(\partial_{5}e^{k}_{\sigma}))
         (A_{\rho}e^{\rho}_{i})(\partial_{5}e^{i}_{\nu})e^{\nu}_{k} \nonumber\\
     & + b\kappa^2(A_{\mu}e^{\mu}_{k})(A_{\rho}e^{\rho}_{i})(g^{\nu\sigma}
         (\partial_{5}e^{i}_{\nu})(\partial_{5}e^{k}_{\sigma})) \nonumber\\
     & + 2c\kappa T^{j}{}_{j}{}^{\sigma}
         (A_{\rho}e^{\rho}_{k})(\partial_{5}e^{k}_{\sigma})
       - 2c\kappa T^{j}{}_{j}{}^{\rho}A_{\rho}e^{\sigma}_{k}
         (\partial_{5}e^{k}_{\sigma}) \nonumber\\
     & + c\kappa^2(A_{\mu}e^{\mu}_{j})(A_{\rho}e^{\rho}_{k})(g^{\nu\sigma}
         (\partial_{5}e^{j}_{\nu})(\partial_{5}e^{k}_{\sigma})) \nonumber\\
     & - 2c\kappa^2(A_{\mu}e^{\mu}_{j})(\partial_{5}e^{k}_{\sigma})e^{\sigma}_{k}
         (g^{\nu\rho}A_{\rho}(\partial_{5}e^{j}_{\nu})) \nonumber\\
     & + c\kappa^2(g^{\mu\rho}A_{\mu}A_{\rho})(\partial_{5}e^{j}_{\nu})e^{\nu}_{j}
         (\partial_{5}e^{k}_{\sigma})e^{\sigma}_{k}\,,\\
\bar{T}_{i\hfive j} \bar{T}^{i\hfive j}
=\,  & \frac{1}{\phi^2}\,\eta^{\hfive\hfive}\eta_{ik}g^{\mu\nu}
         (\partial_{5}e^{i}_{\mu})(\partial_{5}e^{k}_{\nu})\,, \\
\bar{T}_{\hfive ij} \bar{T}^{\hfive ij}
=\,  & \frac{\kappa^2}{4}\phi^2\eta_{\hfive\hfive}
         g^{\mu\rho}g^{\nu\sigma}F_{\mu\nu}F_{\rho\sigma}
       + 2\kappa^3\phi^2\eta_{\hfive\hfive}g^{\mu\sigma}g^{\nu\rho}A_{\mu}
         (\partial_{5}A_{\nu})F_{\sigma\rho} \nonumber\\
     & + 2\kappa^4\phi^2\eta_{\hfive\hfive}(g^{\mu\rho}A_{\mu}A_{\rho})
         (g^{\nu\sigma}(\partial_{5}A_{\nu})(\partial_{5}A_{\sigma})) \nonumber\\
     & - 2\kappa^4\phi^2\eta_{\hfive\hfive}(g^{\mu\sigma}A_{\mu}(\partial_{5}A_{\sigma}))
         (g^{\nu\rho}(\partial_{5}A_{\nu})A_{\rho})\,, \\
\bar{T}_{i\hfive j} \bar{T}^{j\hfive i}
=\,  & \frac{1}{\phi^2}\eta^{\hfive\hfive}(\partial_{5}e^{i}_{\mu})
         (\partial_{5}e^{j}_{\nu})e^{\mu}_{j}e^{\nu}_{i}\,, \\
\bar{T}_{\hfive ij} \bar{T}^{ji\hfive}
=\,  & \frac{\kappa}{2}e^{\nu}_{j}g^{\mu\rho}F_{\mu\nu}(\partial_{5}e^{j}_{\rho})
       - \kappa^2(A_{\mu}e^{\mu}_{j})(g^{\nu\rho}(\partial_{5}A_{\nu})
         (\partial_{5}e^{j}_{\rho})) \nonumber\\
     & + \kappa^2(A_{\mu}g^{\mu\rho})(\partial_{5}A_{\nu})
         (\partial_{5}e^{j}_{\rho})e^{\nu}_{j}\,,\\
\bar{T}_{\hfive i\hfive} \bar{T}^{\hfive i\hfive}
=\,  & \frac{1}{\phi^2}\eta_{\hfive\hfive}\eta^{\hfive\hfive}
         (g^{\mu\nu}(\partial_{\mu}\phi)(\partial_{\nu}\phi))
       + \frac{2\kappa}{\phi^2}\eta_{\hfive\hfive}\eta^{\hfive\hfive}
         (\partial_{5}\phi)(g^{\mu\nu}(\partial_{\mu}\phi)A_{\nu}) \nonumber\\
     & + \frac{2\kappa}{\phi}\eta_{\hfive\hfive}\eta^{\hfive\hfive}
         (g^{\mu\nu}(\partial_{\mu}\phi)(\partial_{5}A_{\nu}))
       + \frac{\kappa^2}{\phi^2}\eta_{\hfive\hfive}\eta^{\hfive\hfive}
         (\partial_{5}\phi)^2(g^{\mu\nu}A_{\mu}A_{\nu}) \nonumber\\
     & + \frac{2\kappa^2}{\phi}\eta_{\hfive\hfive}\eta^{\hfive\hfive}
         (\partial_{5}\phi)(g^{\mu\nu}A_{\mu}(\partial_{5}A_{\nu}))
       + \kappa^2\eta_{\hfive\hfive}\eta^{\hfive\hfive}
         (g^{\mu\nu}(\partial_{5}A_{\mu})(\partial_{5}A_{\nu}))\,, \\
\bar{T}^{j}{}_{ji} \bar{T}^{\hfive}{}_{\hfive}{}^{i}
=\,  & - \frac{1}{\phi}\,T^{\rho}(\partial_{\rho}\phi)
       - \frac{1}{\phi}\,T^{\rho}A_{\rho}(\partial_{5}\phi)
       - \kappa\,T^{\rho}(\partial_{5}A_{\rho}) \nonumber\\
     & - \frac{\kappa}{\phi}(A_{\mu}e^{\mu}_{j})
         (g^{\nu\rho}(\partial_{5}e^{j}_{\nu})(\partial_{\rho}\phi))
       + \frac{\kappa}{\phi}(e^{\nu}_{j}(\partial_{5}e^{j}_{\nu}))
         (g^{\mu\rho}A_{\mu}(\partial_{\rho}\phi)) \nonumber\\
     & - \frac{\kappa^2}{\phi}(\partial_{5}\phi)(A_{\mu}e^{\mu}_{j})
         (g^{\nu\rho}A_{\rho}(\partial_{5}e^{j}_{\nu})) 
       + \frac{\kappa^2}{\phi^2}(\partial_{5}\phi)(\partial_{5}e^{j}_{\nu})e^{\nu}_{j}
         (g^{\mu\rho}A_{\mu}A_{\rho}) \nonumber\\
     & - \kappa^2(A_{\mu}e^{\mu}_{j})(g^{\nu\rho}
         (\partial_{5}e^{j}_{\nu})(\partial_{5}A_{\rho}))
       + \kappa^2(\partial_{5}e^{j}_{\nu})e^{\nu}_{j}
         (g^{\mu\rho}A_{\mu}(\partial_{5}A_{\rho}))\,, \\
\bar{T}^{i}{}_{i\hfive} \bar{T}^{j}{}_{j}{}^{\hfive}
=\,  & \frac{1}{\phi^2}\eta^{\hfive\hfive}(\partial_{5}e^{i}_{\mu})e^{\mu}_{i}
         (\partial_{5}e^{j}_{\nu})e^{\nu}_{j}\,,
\end{align}%
\endgroup
where $\eta_{\hfive\hfive} = \eta^{\hfive\hfive}=-1$ and 
$T_{\text{NGR}}$ is defined in (\ref{E:NGR Lagrangian}).
\end{section}


\begin{section}{Conformal Transformation of  Torsion in $D$-dimension}
\label{B:D-dimensional conformal transformation}
The conformal transformations of the torsion tensor and  vector
in the $D$-dimensional spacetime are given by
\begingroup
\allowdisplaybreaks
\begin{subequations}\label{E:Conformal torsion}
\begin{eqnarray}
{}^{(D)}\widetilde{T}^{I}{}_{MN} & = & \Omega\,{}^{(D)}T^{I}{}_{MN} +
(\partial_{M}\Omega)\,e_{N}^{I} - e_{M}^{I}\,(\partial_{N}\Omega)\,,
\label{E:Conformal torsion tensor}\\
{}^{(D)}\widetilde{T}_{M} & = & {}^{(D)}T_{M} + (1-D)\,\Omega^{-1}\,\partial_{M}\Omega\,,
\label{E:Conformal torsion vector}
\end{eqnarray} 
\end{subequations}
\endgroup
respectively, where $I=\hzero,\hone,\ldots,D-1$ label the orthonormal frame 
and $M,N=0,1,\dots,D-1$  the coordinate frame.
According to (\ref{E:Conformal torsion}), the conformal transformation of the 
extended torsion scalar is given by
\begingroup
\allowdisplaybreaks
\begin{align}\label{E:Conformal extended torsion scalar}
{}^{(D)}T^{(\text{ext})} =\,
& \Omega^2\,{}^{(D)}\widetilde{T}^{(\text{ext})} 
  + \bigg(4\,a + 2\,b - (2-2D)\,c\bigg)\,
     \widetilde{g}^{MN}\widetilde{T}_{M}\,(\Omega\,\partial_{N}\Omega) \nonumber\\
& + \bigg(-2(1-D)\,a -(1-D)\,b + (1-D)^2\,c\bigg)\,
     \widetilde{g}^{MN}\partial_{M}\Omega\,\partial_{N}\Omega \,.
\end{align}
\endgroup
\end{section}


\begin{section}{Conformal Gauge Invariance Condition}\label{A:CGIC}
By taking $2a+b+c=0$ for the kinetic term of $\phi$ in (\ref{E:zero mode action}),
the Lagrangian is reduced to 
\begingroup
\allowdisplaybreaks
\begin{align}\label{E:AB1 Lagrangian}
\left.\mathcal{L}_{g}\right|_{2a+b+c=0} 
= \,e\frac{1}{2\kappa_4}\bigg\{
   \phi\, T_{\text{NGR}} 
   - \frac{a\kappa^2}{4}\phi^{3}\,g^{\mu\rho}g^{\nu\sigma}
   F_{\mu\nu}F_{\rho\sigma} 
- 2c\,g^{\mu\nu}\,T_{\mu}\,\partial_{\nu}\phi\bigg\}\,,
\end{align}
\endgroup
while (\ref{E:lambda=2 Conformal effective Lagrangian}) can be read as
\begingroup
\allowdisplaybreaks
\begin{align}\label{E:AB2 Lagrangian}
\left.\mathcal{L}_{g}\right|_{2a+b+c=0} 
=&\,\widetilde{e}\frac{1}{2\kappa_4}\,
    \bigg\{\widetilde{\phi}\,\bigg(a\widetilde{T}_{\rho\mu\nu} \,\widetilde{T}^{\rho\mu\nu} 
  + b\,\widetilde{T}_{\rho\mu\nu} \,\widetilde{T}^{\nu\mu\rho}\bigg)
    \nonumber\\
& - 6\,c\,
    \widetilde{\phi}\,\widetilde{g}^{\mu\nu}\,\partial_{\mu}\omega\,\partial_{\nu}\omega
 - e^{6\omega}\,\frac{a\kappa^2}{4}
    \widetilde{\phi}^{\,3}\,\widetilde{g}^{\mu\rho}\widetilde{g}^{\nu\sigma}
    \widetilde{F}_{\mu\nu}\widetilde{F}_{\rho\sigma} 
 - 6\,c\,
    \widetilde{g}^{\mu\nu}\,\partial_{\mu}\omega\,\partial_{\nu}\widetilde{\phi}\bigg\}\,.
\end{align}
\endgroup
By comparing (\ref{E:AB1 Lagrangian}),
it is apparent that (\ref{E:AB2 Lagrangian}) 
without the kinetic term and the interaction of $\omega$ and $\phi$ is conformal invariant 
by imposing $c=0$ in the absence of the $A_{\mu}$ field. 
This case coincides with the discussion in Sec.~\ref{Subsec:Conformal Invariant},
but it is not a gauge theory.

If $(2a+b+c)>0$,
we assume that $c/(2a+b+c)=k$, 
$i.e.$, $2a+b=(c/k)-c$.
The Lagrangian (\ref{E:zero mode action}) can be rewritten as 
\begingroup
\allowdisplaybreaks
\begin{align}\label{E:AB3 Lagrangian}
\mathcal{L}_{g} 
=& \,e\frac{1}{2\kappa_4}\bigg\{
   \phi\, (T_{\text{NGR}} 
   - kc\,g^{\mu\nu}\,T_{\mu}T_{\nu})
   - \frac{a\kappa^2}{4}\phi^{3}\,g^{\mu\rho}g^{\nu\sigma}
   F_{\mu\nu}F_{\rho\sigma} 
\nonumber\\
 & + \frac{c}{k\phi}\,\bigg(g^{\mu\nu}
   (\partial_{\mu}-kT_{\mu})\phi
   (\partial_{\nu}-kT_{\nu})\phi\bigg)\bigg\}\,,
\end{align}
\endgroup
where $c\neq0$.
Furthermore, the coefficients in (\ref{E:lambda=2 Conformal effective Lagrangian}) can be read as
\begingroup
\allowdisplaybreaks
\begin{subequations}
\begin{align}
4\,a+2\,b+2\,c &= \frac{2}{k}c\,, \\
14\,a+7\,b+c   &= \bigg(\frac{7}{k}-6\bigg)c\,, \\
2a+b+c         &= \frac{c}{k}\,, \\
8\,a+4\,b-2\,c &= \bigg(\frac{4}{k}-6\bigg)c\,.
\end{align}
\end{subequations}
\endgroup
In order to compare the Lagrangian 
(\ref{E:lambda=2 Conformal effective Lagrangian})
with (\ref{E:AB3 Lagrangian}),
we absorb the term of 
$\widetilde{g}^{\mu\nu}\,\widetilde{T}_{\mu}\,\partial_{\nu}\widetilde{\phi}$ 
into the kinetic term of $\widetilde{\phi}$
in (\ref{E:lambda=2 Conformal effective Lagrangian}).
As a result, (\ref{E:lambda=2 Conformal effective Lagrangian}) becomes
\begingroup
\allowdisplaybreaks
\begin{align}
\mathcal{L}_{g} 
= &   \,\widetilde{e}\frac{1}{2\kappa_4}\,\bigg\{ 
      \widetilde{\phi}\, \widetilde{T}_{\text{NGR}} 
    + \frac{2}{k}c\,
      \widetilde{\phi}\,\widetilde{g}^{\mu\nu}\,\widetilde{T}_{\mu}\,\partial_{\nu}\omega
      \nonumber\\
  & + \bigg(\frac{7}{k}-6\bigg)c\,
      \widetilde{\phi}\,\widetilde{g}^{\mu\nu}\,\partial_{\mu}\omega\,\partial_{\nu}\omega
      \nonumber\\
  & - e^{6\omega}\,\frac{a\kappa^2}{4}
      \widetilde{\phi}^{\,3}\,\widetilde{g}^{\mu\rho}\widetilde{g}^{\nu\sigma}
      \widetilde{F}_{\mu\nu}\widetilde{F}_{\rho\sigma} 
      \nonumber\\
  & + \frac{c}{k\widetilde{\phi}}\,
      \bigg(\widetilde{g}^{\mu\nu}\partial_{\mu}\widetilde{\phi}\,\partial_{\nu}\widetilde{\phi}
    - 2\,k\,\widetilde{g}^{\mu\nu}\,\widetilde{T}_{\mu}\widetilde{\phi}\,\partial_{\nu}\widetilde{\phi}\bigg)
      \nonumber\\
  & + \bigg(\frac{4}{k}-6\bigg)c\,
      \widetilde{g}^{\mu\nu}\,\partial_{\mu}\omega\,
      \partial_{\nu}\widetilde{\phi}\bigg\}\,.
\end{align}
\endgroup
We can use   
$\partial_{\nu}\widetilde{\phi}
      - k\widetilde{T}_{\nu}\widetilde{\phi}
      + k\widetilde{T}_{\nu}\widetilde{\phi}$
instead of 
      $\partial_{\nu}\widetilde{\phi}$, leading to 
\begingroup
\allowdisplaybreaks
\begin{align}
\mathcal{L}_{g} 
= &   \,\widetilde{e}\frac{1}{2\kappa_4}\,\bigg\{ 
      \widetilde{\phi}\, \widetilde{T}_{\text{NGR}} 
    - kc\,\widetilde{\phi}\,\widetilde{g}^{\mu\nu}\,\widetilde{T}_{\mu}\widetilde{T}_{\nu}
      \nonumber\\
  & + \bigg(4-6k+\frac{2}{k}\bigg)c\,
      \widetilde{\phi}\,\widetilde{g}^{\mu\nu}\,\widetilde{T}_{\mu}\,\partial_{\nu}\omega
      \nonumber\\
  & + \bigg(\frac{7}{k}-6\bigg)c\,
      \widetilde{\phi}\,\widetilde{g}^{\mu\nu}\,\partial_{\mu}\omega\,\partial_{\nu}\omega
      \nonumber\\
  & - e^{6\omega}\,\frac{a\kappa^2}{4}
      \widetilde{\phi}^{\,3}\,\widetilde{g}^{\mu\rho}\widetilde{g}^{\nu\sigma}
      \widetilde{F}_{\mu\nu}\widetilde{F}_{\rho\sigma} 
      \nonumber\\
  & + \frac{c}{k\widetilde{\phi}}\,
      \bigg(\widetilde{g}^{\mu\nu}
      (\partial_{\mu}-k\widetilde{T}_{\mu})\widetilde{\phi}\,
      (\partial_{\nu}-k\widetilde{T}_{\nu})\widetilde{\phi}\bigg)
      \nonumber\\
  & + \bigg(\frac{4}{k}-6\bigg)c\,
      \widetilde{g}^{\mu\nu}\,\partial_{\mu}\omega\,
      (\partial_{\nu} - k\widetilde{T}_{\nu})\widetilde{\phi}\bigg\}\,.
\end{align}
\endgroup
We also absorb the term of
$\widetilde{g}^{\mu\nu}\,\partial_{\mu}\omega\,
      (\partial_{\nu} - k\widetilde{T}_{\nu})\widetilde{\phi}$
into the kinetic term of $\tilde{\phi}$ as 
\begingroup
\allowdisplaybreaks
\begin{align}
\mathcal{L}_{g} 
= &   \,\widetilde{e}\frac{1}{2\kappa_4}\,\bigg\{ 
      \widetilde{\phi}\, \widetilde{T}_{\text{NGR}} 
    - kc\,\widetilde{\phi}\,\widetilde{g}^{\mu\nu}\,\widetilde{T}_{\mu}\widetilde{T}_{\nu}
      \nonumber\\
  & + \bigg(\frac{1}{k}+2-3k\bigg)2c\,
      \widetilde{\phi}\,\widetilde{g}^{\mu\nu}\,\widetilde{T}_{\mu}\,\partial_{\nu}\omega
      \nonumber\\
  & + \bigg(\frac{1}{k}+2-3k\bigg)3c\,
      \widetilde{\phi}\,\widetilde{g}^{\mu\nu}\,\partial_{\mu}\omega\,\partial_{\nu}\omega
      \nonumber\\
  & - e^{6\omega}\,\frac{a\kappa^2}{4}
      \widetilde{\phi}^{\,3}\,\widetilde{g}^{\mu\rho}\widetilde{g}^{\nu\sigma}
      \widetilde{F}_{\mu\nu}\widetilde{F}_{\rho\sigma} 
      \nonumber\\
  & + \frac{c}{k\widetilde{\phi}}\,
      \bigg[\widetilde{g}^{\mu\nu}
      \bigg(\partial_{\mu}-k\widetilde{T}_{\mu}
            +(2-3k)\partial_{\mu}\omega\bigg)\widetilde{\phi}\,
      \bigg(\partial_{\nu}-k\widetilde{T}_{\nu}
            +(2-3k)\partial_{\nu}\omega\bigg)\widetilde{\phi}\bigg]\bigg\}\,,\label{E:AB4 Lagrangian}
\end{align}
\endgroup
 so that
\begin{align}
\bigg(\partial_{\mu}-k\widetilde{T}_{\mu}
+(2-3k)\partial_{\mu}\omega\bigg)\widetilde{\phi}
=\, e^{-2\omega}\bigg(\partial_{\mu}-kT_{\mu}\bigg)\phi
\end{align}
transforms as $\phi$ and gives an invariant kinetic term to be
\begin{align}
&\,\widetilde{e}\frac{1}{2\kappa_4}\frac{c}{k\widetilde{\phi}}\,
      \bigg[\widetilde{g}^{\mu\nu}
      \bigg(\partial_{\mu}-k\widetilde{T}_{\mu}
            +(2-3k)\partial_{\mu}\omega\bigg)\widetilde{\phi}\,
      \bigg(\partial_{\nu}-k\widetilde{T}_{\nu}
            +(2-3k)\partial_{\nu}\omega\bigg)\widetilde{\phi}\bigg]
  \nonumber\\
=& \,e\frac{1}{2\kappa_4}\frac{c}{k\phi}\,
      \bigg(g^{\mu\nu}
      (\partial_{\mu}-kT_{\mu})\phi\,
      (\partial_{\nu}-kT_{\nu})\phi\bigg)\,.
\end{align}
In order to have the conformal gauge invariance 
without the vector field $A_{\mu}$ in (\ref{E:AB4 Lagrangian}),
the terms
$\widetilde{\phi}\,\widetilde{g}^{\mu\nu}\,\widetilde{T}_{\mu}\,\partial_{\nu}\omega$ 
and
$\widetilde{\phi}\,\widetilde{g}^{\mu\nu}\,\partial_{\mu}\omega\,\partial_{\nu}\omega$
in (\ref{E:AB4 Lagrangian}) need to be absence, leading to the necessary condition
\begin{equation}\label{E:con gauge necessary condition}
\frac{1}{k}+2-3k=0\,.
\end{equation}
The solutions for (\ref{E:con gauge necessary condition}) are $k=-1/3$ and 1,
resulting in
\begin{equation}
2a+b = -4c
\end{equation}
or
\begin{equation}
2a+b = 0\,.
\end{equation}
Consequently, 
 the corresponding gauge invariant Lagrangian is
\begingroup
\allowdisplaybreaks
\begin{align}
\left.\mathcal{L}_{g}\right|_{k=-\frac{1}{3}} 
=&    \,\widetilde{e}\frac{1}{2\kappa_4}\,\bigg\{ 
      \widetilde{\phi}\, \widetilde{T}_{\text{NGR}}  
    + \frac{1}{3}c\,\widetilde{\phi}\,\widetilde{g}^{\mu\nu}\,\widetilde{T}_{\mu}\widetilde{T}_{\nu}
      \nonumber\\
  & - \frac{3c}{\widetilde{\phi}}\,
      \bigg[\widetilde{g}^{\mu\nu}
      \bigg(\partial_{\mu} 
        + \frac{1}{3}\widetilde{T}_{\mu}
        + 3\partial_{\mu}\omega\bigg)\widetilde{\phi}\,
      \bigg(\partial_{\nu} 
        + \frac{1}{3}\widetilde{T}_{\nu}
        + 3\partial_{\nu}\omega\bigg)\widetilde{\phi}\bigg]\bigg\}\,
\end{align}
\endgroup
for  $k=-1/3$ and 
\begingroup
\allowdisplaybreaks
\begin{align}
\left.\mathcal{L}_{g}\right|_{k=1} 
= &   \,\widetilde{e}\frac{1}{2\kappa_4}\,\bigg\{ 
      \widetilde{\phi}\bigg(a\widetilde{T}_{\rho\mu\nu} \,\widetilde{T}^{\rho\mu\nu} 
    - 2a\widetilde{T}_{\rho\mu\nu} \,\widetilde{T}^{\nu\mu\rho}\bigg) 
      \nonumber\\
  & + \frac{c}{\widetilde{\phi}}\,
      \bigg(\widetilde{g}^{\mu\nu}
      (\partial_{\mu} - \widetilde{T}_{\mu} - \partial_{\mu}\omega)\widetilde{\phi}\,
      (\partial_{\nu} - \widetilde{T}_{\nu} - \partial_{\nu}\omega)\widetilde{\phi}\bigg)\bigg\}\,
\end{align}
\endgroup
for $k=1$.

\end{section}

\end{document}